\documentclass[10pt,final,letter,twocolumn,twoside,journal,romanappendices]{IEEEtran}
 \usepackage{cite}
\usepackage{epstopdf}
\usepackage{epsfig}
\usepackage{amsthm}
\usepackage{amsmath,amssymb,amsfonts,amstext,amsbsy,amsopn,dsfont}
\usepackage{cases}
\usepackage{bigints}
\usepackage{sublabel}
\usepackage{xcolor}
\usepackage{float}
\usepackage{array}
\usepackage{cuted}

\newtheorem{theorem}{\textbf{Theorem}}
\newtheorem{proposition}{\textbf{Proposition}}

\newcommand{\defn}{\triangleq}
\newcommand{\dif}{\mathrm{d}}

\begin{document}

\title{A 3D Tractable Model for UAV-Enabled Cellular Networks With Multiple Antennas}

\author{Chun-Hung Liu, \IEEEmembership{Senior Member, IEEE}, Di-Chun Liang, Md Asif Syed, and Rung-Hung Gau  
\thanks{C.-H. Liu and M. A. Syed are with the Department of Electrical and Computer Engineering, Mississippi State University, MS 39762, USA. (e-mail: chliu@ece.msstate.edu; ms4273@msstate.edu)}
\thanks{D.-C. Liang and R.-H. Gau are with the Institute of Communications Engineering and Department of Electrical and Computer Engineering, National Chiao Tung University, Hsinchu 30010, Taiwan. (e-mail: ldc30108@gmail.com; runghunggau@g2.nctu.edu.tw)}}	


\maketitle

\begin{abstract}
 This paper aims to propose a three-dimensional (3D) point process that can be employed to generally deploy unmanned aerial vehicles (UAVs) in a large-scale 3D cellular network and to tractably analyze the fundamental network-wide performances of the network. The proposed 3D point process is devised based on a 2D marked Poisson point process in which each point and its random mark uniquely correspond to the projection and the altitude of each point in the 3D point process, respectively. We study some of the important statistical properties of the proposed 3D point process and shed light on some crucial insights into them that facilitate the analyses of a UAV-enabled cellular network wherein all UAVs equipped with multiple antennas are deployed by the proposed 3D point process to serve as aerial base stations. The salient features of the proposed 3D point process lie in its suitability in practical 3D channel modeling and tractability in analysis. The downlink coverages of the UAV-enabled cellular network are found and their closed-form results for some special cases are also derived. Most importantly, their fundamental limits achieved by cell-free massive antenna array are characterized when coordinating all the UAVs to jointly perform non-coherent downlink transmission. These key findings and observations are numerically validated in this paper.
\end{abstract}


\vspace{-0.1in}
 
\section{Introduction}
\IEEEPARstart{I}{n} recent years, the technology of unmanned aerial vehicle (UAV) has been improved significantly such that UAVs possess an outstanding capability of agilely moving in three-dimensional (3D) space, which has attracted increasing attention from the academia and industry of wireless communications because such a 3D moving capability is able to remarkably relieve spatial limitations, which usually lead to the impairments of wireless channels between two static terminals, such as path loss, penetration loss, and multi-path fading, etc. Qualcomm and AT\&T, for example, have been planning to build up a \textit{UAV-enabled cellular network} with UAVs working as ``aerial base stations'' in order to enable large-scale wireless communications in the upcoming fifth generation (5G) cellular networks~\cite{Qualcomm18}. Moreover, Amazon prime air and Google’s drone delivery project are two striking examples of using ``cellular-connected'' UAV communications where UAVs are aerial mobile users in a cellular network~\cite{DS17}. Although the agile and flexible mobility of UAVs benefits point-to-point communications between UAVs and other terminals, it may not really facilitate communications in a wireless network where many UAVs are arbitrarily  deployed and a considerable amount of co-channel interference is created accordingly. As such, how to appropriately deploy UAVs in a wireless network to reap the mobility advantage of UAVs is a prominent problem pertaining to all the aspects of UAV communications and networking. 

The 3D deploying problem for a UAV-enabled cellular network with UAVs serving as aerial base stations is involved in the issue of simultaneous multi-user coverage, and thereby it is much more complicated and difficult than the 3D deploying problem for a cellular-connected UAV network that merely needs to tackle the issue of single UAV coverage at a time. Deploying methods for a UAV-enabled cellular network should be able to exploit the mobility of UAVs in order to ameliorate the fundamental coverage limit of the entire cellular network, yet how to evaluate the deploying methods in a \textit{tractable} and \textit{network-wide} way remains unclear until now. The key to tackling the problems of deploying and evaluating a UAV-enabled cellular network lies in the tractability of the modeling framework of UAV-enabled cellular networks that certainly depends upon the randomly distributed nature of UAVs hovering in the sky. On account of this, we propose a 3D deployment model for a UAV-enabled cellular network, which is able to not only generally characterize the spatial random distribution of the UAVs in the network, but also skillfully pave a tractable way to analyze the performances of the UAV-enabled cellular network. The 3D deployment model proposed to deploy a UAV-enabled cellular network is devised based on a 3D point process, which is essentially a 2D homogeneous \textit{marked} Poisson point process (PPP) in that all the projections and the altitudes of the 3D point process respectively consist of all the points and the marks of the 2D homogeneous marked PPP. Such a 3D point process is able to generally and practically characterize the randomly positioning characteristic of UAVs in a large-scale UAV-enabled cellular network so that it is very  distinct from the existing UAV-related deployment models in the literature, as reviewed in the following. 

\subsection{Prior Works on Modeling UAV-Enabled Wireless Networks}
Many of the prior works on UAV-enabled cellular networks studied their problems by assuming a fixed number of UAVs deployed in the sky (typically see \cite{AMMRFCKCPS17,YSDYTXDYZJ19,WZDLZR19,ZYZR17,OFOHMR16,YXFZXWNWZJAW18}).  References \cite{AMMRFCKCPS17,YSDYTXDYZJ19,WZDLZR19}, for example, adopted a single UAV in a wireless network to analyze the performance metrics of the network, such as outage probability, energy efficiency, and throughput, to see how to position a UAV so that the performance metrics can be maximized. The problem of how to jointly optimize the flight radius and speed of a single UAV so as to maximize the energy efficiency of a UAV was tackled in~\cite{ZYZR17}, whereas the problem of how to use a variable-rate relaying approach to optimizing outage probability and information rate for a single UAV was studied in~\cite{OFOHMR16}. Reference \cite{YXFZXWNWZJAW18} analyzed the link capacity for two UAVs with random 3D trajectories and then addressed the impacts of network densification, imperfect channel state information, and interference on the link capacity. Although the modeling and analysis approaches of these prior works seem suitable for a wireless network with a small number of UAVs, in general they cannot be straightforwardly employed in large-scale UAV-enabled wireless networks, which need to take into account of the interactions between UAVs.

There are indeed some prior works that modeled UAV-enabled wireless networks in a large-scale sense, e.g., \cite{VVCHSD17,HTLYSZSXCY1902,SYDZDX19,KDLJQTQ19,AMMGGGRAPS20,YJXJ19,GBKJDLA17}. However, the majority of them simply assumed that all UAVs in a network hover at the same \textit{fixed altitude}. For example, reference \cite{VVCHSD17} investigated the coverage problem for a finite network model assuming a number of UAVs are uniformly distributed at the same fixed altitude in the network. The coverages based on UAV-centric and user-centric strategies for multi-UAV-assisted NOMA networks were studied in~\cite{HTLYSZSXCY1902}. Reference~\cite{SYDZDX19} proposed  a UAV-assisted wireless network for the malfunction areas and used a user-centric cooperation scheme to evaluate the coverage and normalized spectral efficiency of the network. A multi-layer UAV network was proposed in~\cite{KDLJQTQ19}  to analyze and optimize the successful transmission probability and spectral efficiency of the network, while the coverage and ergodic rate of a UAV-enabled network were investigated with a spectrum sharing mechanism in~\cite{AMMGGGRAPS20}. These prior works all assumed that all UAVs hover at the same fixed altitude in a network so that their analyses cannot practically reflect how they are influenced by a real-world deployment of UAVs with a \textit{random} altitude.  

Some prior works already tried to relax the modeling assumption of ``fixed altitude'' when modeling multiple UAVs in the sky. Reference \cite{SBLLRRMMVRJH20}, for example, studied the coverage probability in a 3D deployment model of UAVs wherein all UAVs were distributed within a specific range of altitude that was uniformly divided into a certain number of levels. Reference \cite{SBLLRRMMVRJH20} considered that UAVs were uniformly distributed 
above a 2D plane and positioned at different levels of altitude. A few prior works also adopted 3D homogeneous PPPs to model UAV-enabled cellular networks.  Reference~\cite{CHKHHJYW19} exploited the limits of the coverage and volume spectral efficiency of a mmWave UAV cellular network in which a UAV's altitude was modeled as a function of the UAV's projection. The coverage and network throughput of a NOMA-assisted UAV network modeled by a 3D homogeneous PPP were analyzed in~\cite{HTLYSZSXCY2003}, whereas reference~\cite{ZCZW16} considered spectrum sharing when analyzing the success probability and total network throughput of a UAV-enabled network modeled by a 3D PPP.  Modeling the distribution of UAVs by 3D PPPs entails two practical issues. One is that UAVs are low-altitude platforms and cannot be arbitrarily positioned in infinitely large 3D space modeled by a 3D PPP. The other is that the path-loss exponent of any wireless links in a wireless network modeled by a 3D PPP needs to be greater than three in order to make analysis bounded, yet such a constraint on the path-loss exponent is not practically true for most 3D wireless links with a path-loss exponent smaller than three. 

\subsection{Contributions}
Although these aforementioned prior works successfully conducted some analyses for specific problems they are interested in, in general their outcomes are not easily generalized to a network-wide scenario in a large-scale UAV-enabled cellular network in that their generality is subject to their simplified models and assumptions of deploying UAVs in a wireless network. Our proposed 3D deployment model for a UAV-enabled cellular network, as will be shown in the following sections, inherits the tractability of employing PPPs to model and analyze a wireless network, offers an additional degree of freedom in controlling the altitude of a UAV, and more importantly fits practical 3D path-loss channel models. Consequently, the analytical results of this paper are more general and closer to practical results in a UAV-enabled cellular network.  The main contributions of this paper are summarized as follows.
\begin{itemize}
	\item A 3D point process is proposed based on a 2D homogeneous \textit{marked} PPP in which each point and its mark are the terrestrial  projection and the random altitude of a unique point in the 3D point process, respectively. This 3D point process is shown to work for all practical 3D path-loss channel models between any two points in the point process. 
	\item In the proposed 3D point process, we consider \textit{angle-projection-independent locating} (APIL) and \textit{angle-projection-dependent locating} (APDL) scenarios when positioning all the UAVs in the sky. The APIL scenario refers to when the elevation angle and the projection of each UAV are independent, whereas the APDL scenario refers to the opposite. The proposed 3D point process is shown to be essentially a 2D homogeneous PPP in the APIL scenario, yet it is shown to be equivalent to a 2D non-homogeneous PPP in the APDL scenario.
	\item The fundamental properties of the proposed 3D point process are analyzed for the APIL and APDL scenarios, which facilitate the derivations and analyses of the Laplace transforms of the \textit{complete and truncated} (incomplete) \textit{3D shot signal processes} with considering line-of-sight (LoS) and non-line-of-sight (NLoS) channel behaviors.   
	\item The proposed 3D point process is employed to generally model the random deployment of the UAVs that are equipped with multiple antennas and serve as aerial base stations in a large-scale cellular network and the \textit{downlink coverages} (probabilities) for the APIL and APDL scenarios are explicitly found and some of them are shown to reduce to a closed-form expression for special channel conditions.
	\item The \textit{cell-free downlink coverages} for the APIL and APDL scenarios are explicitly derived when all the UAVs in the network can do non-coherent joint transmission. They represent the fundamental upper limits of the downlink coverage probabilities that are achievable in the two scenarios. Their closed-form expressions for some special channel condition are obtained as well.  
\end{itemize}
Furthermore, we provide numerical results to validate the correctness of the analytical findings of the downlink coverages in this paper and show that in general the downlink coverages are  insensitive to the different distributions of the elevation angle and the altitude of UAVs that have the same mean so that they can be approximated by the derived expressions using the mean of the elevation angle of a UAV for APIL and the mean of the altitude of a UAV for APDL.

\subsection{Paper Organization}
The rest of this paper is organized as follows. In Section~\ref{Sec:SystemModel}, a 3D point process is proposed and some of its important statistical properties are studied. We employ the proposed 3D point process to model a 3D UAV-enabled cellular network consisting of a tier of UAVs serving as aerial base stations with multiple antennas and we then analyze the downlink coverage performances of the UAV-enabled cellular network in Section~\ref{Sec:UAVNetworkModel}. Section~\ref{Sec:Simulation} provides some numerical results in order to validate the analytical findings in Section~\ref{Sec:UAVNetworkModel}. Finally, Section~\ref{Sec:Conclusion} concludes the important findings in this paper.  

\section{The Proposed 3D Point Process and Its Statistical Properties}\label{Sec:SystemModel}

\begin{figure*}[t!]
	\centering
	\includegraphics[width=\textwidth, height=3.0in]{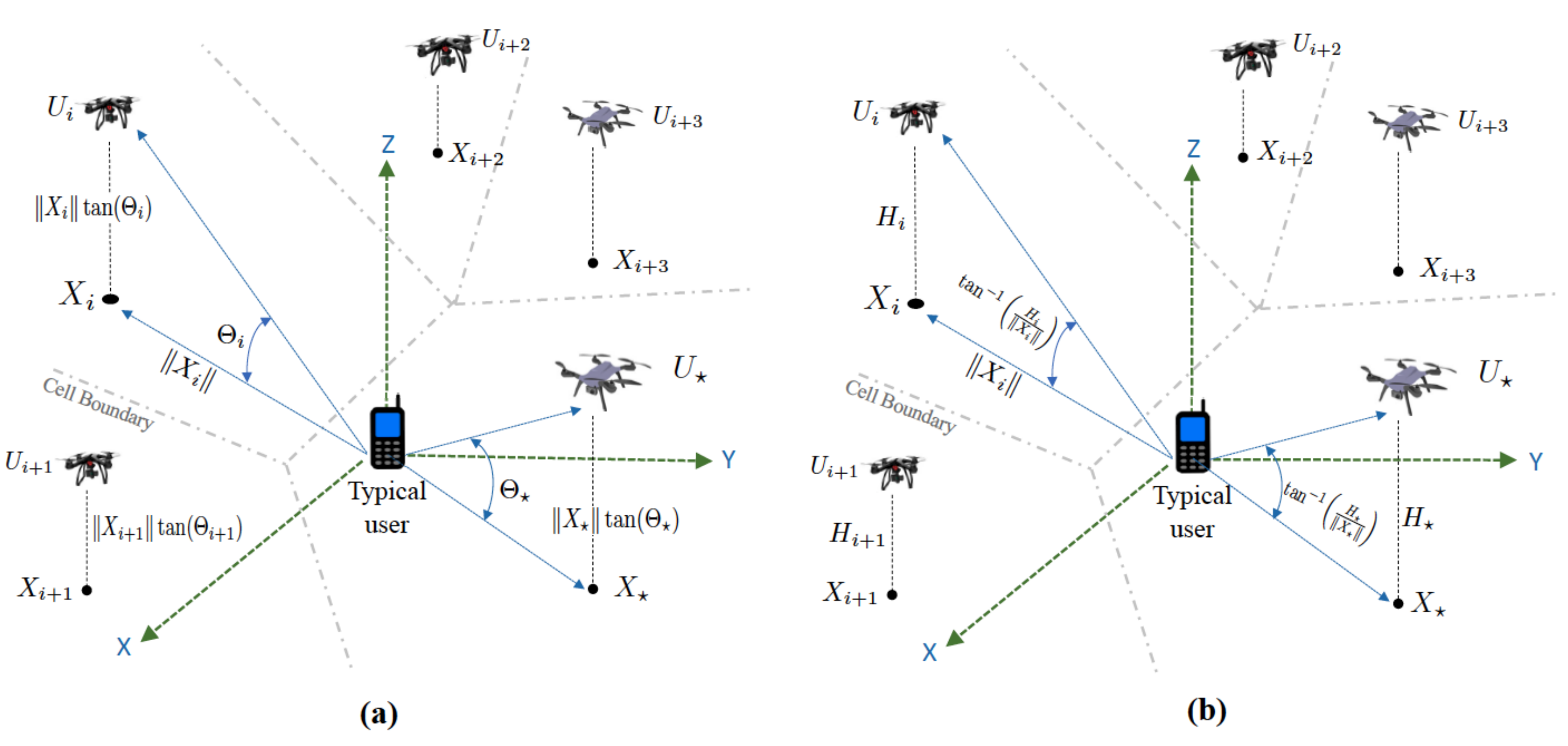}
	\caption{The proposed 3D point process $\Phi_u$ is used to model the locations of the UAVs in a cellular network and the projections of all the points in $\Phi_u$ form a 2D homogeneous PPP of density $\lambda$ on the $\mathsf{X}-\mathsf{Y}$ plane. The projection of point $U_i$ is denoted by $X_i$ and a typical user located at the origin associates with UAV $U_{\star}$ serving as its aerial base station. Two scenarios of locating the UAVs are considered: (a) The APIL scenario: $\Theta_i$ and $\|X_i\|$ are independent for all $i\in\mathbb{N}_+$ and $H_i=\tan(\Theta_i)\|X_i\|$ depends on $\Theta_i$ and $X_i$. (b) The APDL scenario: $H_i$ and $\|X_i\|$ are independent for all $i\in\mathbb{N}_+$ so that $\Theta_i=\tan^{-1}(H_i/\|X_i\|)$ is dependent upon $H_i$ and $X_i$. }
	\label{Fig:SystemModel}
\end{figure*}

Suppose a 2D homogeneous PPP of density $\lambda$ can be denoted by the following set on the plane of $\mathbb{R}^2$
\begin{align}
\Phi_x\defn \{X_i\in\mathbb{R}^2: i\in\mathbb{N}_+\},
\end{align}
and it is assumed to be a \textit{simple} point process, that is, none of the points in $\Phi_x$ can have the same location on the plane of $\mathbb{R}^2$. In accordance with $\Phi_x$, we propose the following 3D point process $\Phi_u$:
\begin{align}\label{Eqn:3DPP}
\Phi_u \defn \bigg\{ & U_i\in\mathbb{R}^2\times \mathbb{R}_+: U_i=(X_i,H_i), X_i\in\Phi_x,\nonumber\\ 
& H_i=\|X_i\|\tan(\Theta_i), \Theta_i\in\left[0,\frac{\pi}{2}\right], i\in\mathbb{N}_+\bigg\},
\end{align}
where $X_i$ is the projection of point $U_i$ on the plane of $\mathbb{R}^2$, $\|X_i\|$ is the distance between the origin\footnote{Without loss of generality, in this paper we use the origin as a reference point for the locations of the points in point sets such as $\Phi_x$ and $\Phi_u$ to express their relevant equations, results, and observations. According to the Slivnyak theorem~\cite{DSWKJM13}\cite{MH12}, the statistical properties of a PPP evaluated at the origin are the same as those evaluated at any particular point in the PPP.} and $X_i$,  and $\Theta_i$ is the (random) elevation angle from the origin to point $U_i$. Hence, the ``altitude'' of point $U_i$ is $H_i$ that is the distance from $X_i$ to $U_i$ such that $\Phi_u$ can be viewed as a marked version of $\Phi_x$ in which each point has a mark as its altitude. Since $\|Y_i-Y_j\|$ denotes the Euclidean distance between points $Y_i$ and $Y_j$ for $i\neq j$, we know $\|X_i\|=\|U_i\|\cos(\Theta_i)$ and thus $\|U_i\|=\|X_i\|\sec(\Theta_i)$. A link between two spatial points is called a LoS link provided it is not visually blocked from one point to the other. A low-altitude-platform communication scenario is considered in this paper and the LoS model of a 3D channel in~\cite{AHKSSL14} is adopted so that we have the following LoS probability of the 3D channel between the origin and a point $U_i\in\Phi_u$ proposed in~\cite{AHKSSL14}:
\begin{align}\label{Eqn:LoSProb}
\rho\left(\Theta_i\right) \defn \frac{1}{1+c_2\exp\left(-c_1\Theta_i\right)},
\end{align}
where $c_1$ and $c_2$ are environment-related positive constants (for rural, urban, etc.), and thereby whether or not point $U_i$ is LoS for the origin is completely determined by its elevation angle $\Theta_i$ from the origin. 

For the 3D point process $\Phi_u$,  we will specifically consider two positioning scenarios for the points in $\Phi_u$, i.e., the \textit{angle-projection-independent locating} (APIL) and the \textit{angle-projection-dependent locating} (APDL) scenarios\footnote{Note that the APDL scenario can also be referred to as the \textit{altitude}(\textit{height})-projection-independent positioning scenario because the APDL scenario is essentially defined in a way that the altitude and the projection of a point in $\Phi_u$ are independent.}. An illustration of using the proposed 3D point process $\Phi_u$ to deploy UAVs based on these two scenarios is depicted in Fig.~\ref{Fig:SystemModel}. In the figure, the projections of the UAVs on the $\mathsf{X}-\mathsf{Y}$ (ground) plane form a 2D homogeneous PPP $\Phi_x$. In the APIL scenario, as shown in part (a) of the figure, the elevation angle and the projection of each point in $\Phi_u$ are independent, that is, $\Theta_i$ and $X_i$ are independent for all $i\in\mathbb{N}_+$, and we thus can use $\Theta_i$ and $X_i$ to completely define the 3D location of UAV $U_i$. In contract, part (b) of the figure depicts the APDL scenario in which the altitude and the projection of a UAV are independent so that they can also be adopted to completely describe the 3D location of a UAV. Note that the elevation angle and the projection of a UAV is no longer independent once the altitude and the projection of the UAV are independent because they can be used to find the elevation angle between them, i.e., $\Theta_i=\tan^{-1}(H_i/\|X_i\|)$for all $i\in\mathbb{N}$. In fact, the APIL scenario corresponds the scenario of employing the 3D polar coordinate system to describe the locations of UAVs, whereas the APDL corresponds to the scenario of employing the Cartesian coordinate system to describe the locations of UAVs. The main motivation of considering the APDL and APDL scenarios in this paper is inspired by the fact that they both have their practical deployment applications. The APIL scenario properly characterizes the deployment situation that when each UAV in $\Phi_u$ is positioned at a random ground point and at a random elevation angle whose distribution can be observed at the origin. Such a situation usually happens when there is a ground central controller that adjusts the elevation angle of a UAV so as to make the wireless channel between the UAV and a user more likely to be an LoS one. The APDL scenario appropriately characterizes the situation that when each UAV in $\Phi_u$ is positioned at a random projection and at a random altitude whose distribution is known. Such a situation frequently occurs when each point (UAV) in $\Phi_i$ is positioned by a ground central controller at a random altitude in accordance with some specific rule. In the following, we will analyze some important statistical properties related to $\Phi_u$ by considering these two scenarios.

\subsection{Distance-Related Distributions in $\Phi_u$}\label{SubSec:Dis-RelDis}
Suppose a non-negative RV $R_{\star}$ is defined as
\begin{align}\label{Eqn:Weight3DDis}
R_{\star}\defn \max_{i:U_i\in\Phi_u} \left\{W_iL_i\|U_i\|^{-\alpha}\right\},
\end{align}
where $\alpha>2$ is a constant\footnote{If $\|U_i\|^{-\alpha}$ stands for the path loss between node $U_i$ and the origin, $\alpha$ is referred to as the path-loss exponent, which will be used in Section~\ref{Sec:UAVNetworkModel}. Moreover, the channel model adopted in~\ref{Eqn:Weight3DDis} is for wireless channels in the UHF band. For wireless channels in much higher frequency bands (e.g., the mmWave band), a much complicated channel model should be adopted to properly characterize the NLoS effects on the channels, such as the channel models adopted in~\cite{CHKHHJYW19,WYYLYD20}.}, $L_i\in\{1,\ell\}$ is a Bernoulli RV that is equal to one if an LoS link between the origin and point $U_i$ exists and $\ell$ otherwise, and $W_i\in\mathbb{R}_+$ is a non-negative weighting RV associating with $U_i$ and independent of all $L_i$'s and $U_i$'s. Note that $\ell\in[0,1]$ is referred to as the NLoS channel attenuation factor since it is used to model the penetration loss of an NLoS link, $W_i$ is independent of $L_j$ and $U_j$ for all $i,j\in\mathbb{N}_+$, all $W_i$'s are assumed to be independently and identically distributed (i.i.d.), and the distribution of $L_i$ depends on the location of $U_i$ as indicated by the LoS probability in~\eqref{Eqn:LoSProb}. Throughout this paper, all the evaluation angles in $\Phi_u$ are assumed to be i.i.d. for the APIL scenario and all the altitudes in $\Phi_u$ are i.i.d. for the APDL scenario. We then have the following theorem.
 
\begin{theorem}\label{Thm:CDFR*}
Suppose the moment of $W_i$ exists (i.e., $\mathbb{E}[W_i^{a}]<\infty$ for all $a>0$) for all $i\in\mathbb{N}_+$. (i) If the APIL scenario is considered such that $\Theta_i$ and $\|X_j\|$ are independent for all $i,j\in\mathbb{N}_+$ and all $\Theta_i$'s are independently and identically distributed (i.i.d.), the cumulative density function (CDF) of $R_{\star}$ defined in~\eqref{Eqn:Weight3DDis}  can be found as
\begin{align}\label{Eqn:CDFR*1}
F_{R_{\star}}(r) = &\exp\left(-\pi\lambda \mathbb{E}\left[W^{\frac{2}{\alpha}}\right]\omega  r^{-\frac{2}{\alpha}}\right),
\end{align}
where $F_Z(\cdot)$ denotes the CDF of RV $Z$ and $\omega$ is defined as
\begin{align}\label{Eqn:Density3DPPInd}
\omega\defn  \mathbb{E}\left\{\cos^2(\Theta)\left[\rho(\Theta)\left(1-\ell^{\frac{2}{\alpha}}\right) +\ell^{\frac{2}{\alpha}}\right]\right\}.
\end{align}
(ii) If the APDL scenario is considered such that $\Theta_i=\tan^{-1}(H_i/\|X_i\|)$ and all $H_i$'s are i.i.d., then $F_{R_{\star}}(r)$ can be derived as
\begin{align}\label{Eqn:CDFR*2}
F_{R_{\star}}(r) = \exp\left[-\pi \lambda\Omega\left(r\right)\right],
\end{align}
where $\Omega(r)$ is defined as
\begin{align}\label{Eqn:Density3DPPNonInd}
\Omega\left(r\right) \defn  \int_0^{\infty} & \mathbb{E}_H\bigg\{\bigg[\rho\left(\Theta\right) F^c_{W}\left(r(z+H^2)^{\frac{\alpha}{2}}\right)\nonumber\\
&+[1-\rho\left(\Theta\right)]F^c_{W}\left(\frac{r}{\ell}(z+H^2)^{\frac{\alpha}{2}}\right)\bigg]\bigg\} \dif z
\end{align}
in which $\Theta=\tan^{-1}(H/\sqrt{z})$ and $F^c_W(\cdot)$ denotes the complementary CDF (CCDF) of RV $W$.
\end{theorem}
\begin{IEEEproof}
See Appendix \ref{App:ProofCDFR*}.
\end{IEEEproof}
\noindent Note that the expression in~\eqref{Eqn:CDFR*2} is more complicated than its corresponding closed-form expression in~ \eqref{Eqn:CDFR*1} due to the dependence between the elevation angle $\Theta_i$ and the projection $X_i$ of point $U_i$ in $\Phi_u$ for all $i\in\mathbb{N}_+$ and it reduces to $\eqref{Eqn:CDFR*1}$ once the dependence does not exist (Namely, $\Theta$ in~\eqref{Eqn:Density3DPPNonInd} is not a function of $H$ and $z$.). Furthermore, in general $\omega$ in~\eqref{Eqn:Density3DPPInd} is insensitive to the different distributions of $\Theta$ that have the same mean, especially when the mean of $\Theta$ is not very large. This point will be numerically demonstrated in Section~\ref{SubSection:SimAPIL}.

The results in~\eqref{Eqn:CDFR*1} and ~\eqref{Eqn:CDFR*2} are very general since they are valid for the general distributions of $W$ and $\Theta$. Accordingly, they can be employed to find the distributions of some specific RVs related to set $\Phi_u$. To demonstrate this, we discuss some special cases of $R_{\star}$ in the following.
\subsubsection{$W_i=L_i=1$} In this case, $R_{\star}$ in~\eqref{Eqn:Weight3DDis} reduces to $R_{\star}=\max_{U_i\in\Phi_u}\|U_i\|^{-\alpha}$ so that $R^{-\frac{1}{\alpha}}_{\star}=\min_{U_i\in\Phi_u}\|U_i\|$ is the shortest distance between the origin and set $\Phi_u$. Thus, using $F_{R_{\star}}(r)$ in~\eqref{Eqn:CDFR*1} helps find the CCDF of $R^{-\frac{2}{\alpha}}_{\star}$ as
\begin{align}\label{Eqn:CCDFInvR*1}
F^c_{R^{-\frac{2}{\alpha}}_{\star}}(y) = \exp\bigg(-\pi\lambda\mathbb{E}\left[\cos^2(\Theta)\right]  y\bigg),
\end{align}
which indicates that $R^{-2/\alpha}_{\star}\sim\exp(\pi\lambda\mathbb{E}[\cos^2(\Theta)])$ is an exponential RV with mean $1/\pi\lambda\mathbb{E}[\cos^2(\Theta)]$, and it is exactly the CCDF of the square of the shortest distance between the origin and  a 2D homogeneous PPP of density $\lambda \mathbb{E}[\cos^2(\Theta)]$~\cite{DSWKJM13,MH12}. Namely, this observation manifests that \textit{the 3D point process $\Phi_u$ proposed in~\eqref{Eqn:3DPP} can be equivalently viewed as a 2D homogeneous PPP of density $\lambda \mathbb{E}[\cos^2(\Theta)]$ as long as the elevation angle and the projection of each point in $\Phi_u$ are independent}. Moreover, using ~\eqref{Eqn:CDFR*2} for this case yields $F^c_{R^{-2/\alpha}_{\star}}(y)$ given by
\begin{align}\label{Eqn:CCDFInvR*2}
F^c_{R^{-\frac{2}{\alpha}}_{\star}}(y) &=\exp\left[-\pi\lambda\Omega\left(y^{-\frac{\alpha}{2}}\right)\right]\nonumber\\
&=\exp\left[-\pi \int_0^{y} \lambda F_H(\sqrt{z}) \dif z\right].
\end{align}
For this case, $\Phi_u$ can be viewed as a 2D \textit{non-homogeneous} PPP of location-dependent density $\lambda F_H(\sqrt{z})$. Thus, we can conclude that \textit{$\Phi_u$ becomes a 2D non-homogeneous PPP whenever the elevation angle and the projection of each point in $\Phi_u$ are not independent (i.e., the APDL scenario)}. There is a simple example which demonstrates this scenario, that is, if all the points in $\Phi_u$ are positioned at the same altitude of $\overline{h}$, we know  $\Theta_i=\tan^{-1}(\overline{h}/\|X_i\|)$ so that $F^c_{R^{-2}_{\star}}(y)=\exp(-\pi \int_0^{y}\lambda\mathds{1}(\overline{h}\leq\sqrt{z}) \dif z)$ and $\Phi_u$ is a non-homogeneous PPP of density $\lambda \mathds{1}(\overline{h}\leq \sqrt{z})$ for this example. 

\subsubsection{$W_i=1$} For this case, $R_{\star}$ in~\eqref{Eqn:Weight3DDis} reduces to $R_{\star}=\max_{i:U_i\in\Phi_u} L_i\|U_i\|^{-\alpha}$ and thus $R^{-\frac{1}{\alpha}}_{\star}=\min_{i:U_i\in\Phi_u}\{L^{-\frac{1}{\alpha}}_i\|U_i\|\}$.
Thus, the distribution of $R_{\star}^{-\frac{1}{\alpha}}$ can reflect how the LoS effect impacts the distribution of the shortest distance between the origin and set $\Phi_u$. By considering $W=1$ in~\eqref{Eqn:CDFR*1}, we can obtain $F^c_{R^{-\frac{2}{\alpha}}_{\star}}(y)$ as  shown in the following:
\begin{align}\label{Eqn:CCDFInvR*3}
F^c_{R^{-\frac{2}{\alpha}}_{\star}}(y) = \exp\left(-\pi \lambda\omega  y\right),
\end{align} 
i.e., $R^{-\frac{2}{\alpha}}_{\star}\sim\exp(\pi\lambda\omega)$, which reveals that the following point set
\begin{align}
\widetilde{\Phi}_u\defn\left\{\widetilde{U}_i\in\mathbb{R}^2\times\mathbb{R}_+: \widetilde{U}_i=L^{-\frac{1}{\alpha}}_iU_i, L_i\in\{1,\ell\}, U_i\in\Phi_u\right\} \label{Eqn:Equ3DPointProcess}
\end{align}
can be viewed as a thinning PPP from $\Phi_x$ with density $\lambda\omega$. When $\ell=0$, $R^{-\frac{1}{\alpha}}_{\star}$ is the shortest distance of the LoS link from the origin to set $\Phi_u$ and $F^c_{R^{-\frac{2}{\alpha}}_{\star}}(x)$ in~\eqref{Eqn:CCDFInvR*3} reduces to $e^{-\pi\lambda \mathbb{E}[\rho(\Theta)\cos^2(\Theta)]y}$. Therefore,  \textit{in the APIL scenario the LoS points in $\Phi_u$ are equivalent to a 2D homogeneous PPP of density $\lambda \mathbb{E}[\cos^2(\Theta)\rho(\Theta)]$}. Furthermore, for $F^c_{R^{-\frac{2}{\alpha}}_{\star}}(y)$ in~\eqref{Eqn:CDFR*2} with $W=1$, we can have
\begin{align}
F^c_{R^{-\frac{2}{\alpha}}_{\star}}(y) =\exp\left[-\pi\lambda  \Omega\left(y^{-\frac{\alpha}{2}}\right) \right],
\end{align}
where $\Omega(y^{-\alpha/2})$ is found from~\eqref{Eqn:Density3DPPNonInd} for $W=1$ and it is given by
\begin{align}
\Omega\left(y^{-\frac{\alpha}{2}}\right)= \mathbb{E}_H\bigg[ &\left(y\ell^\frac{2}{\alpha}-H^2\right)^++\int^{(y-H^2)^+}_{(y\ell^\frac{2}{\alpha}-H^2)^+}\nonumber\\
&\rho\left(\tan^{-1}\left(\frac{H}{\sqrt{z}}\right)\right)\dif z\bigg], \label{Eqn:OmegaFunLos}
\end{align}
where $(x)^+\defn\max\{0,x\}$. Hence, $\widetilde{\Phi}_u$ can be viewed as a 2D non-homogeneous PPP of density $\lambda\frac{\dif\Omega(z)}{\dif z}$ since $R^{-\frac{1}{\alpha}}_{\star}$ is the shortest distance from the origin to $\widetilde{\Phi}_u$. More specifically, if $\ell=0$ and $H$ is equal to a constant $\overline{h}>0$, we further know $\Omega(y^{-\frac{\alpha}{2}})=\int_0^{(y-\overline{h}^2)^+}\rho(\tan^{-1}(\overline{h}/\sqrt{z}))\dif z$, which reveals that \textit{the LoS points in $\Phi_u$ can be equivalently viewed as a 2D non-homogeneous PPP of density $\lambda\rho(\tan^{-1}(\overline{h}/\sqrt{z}))$ in the APDL scenario}. Also, \eqref{Eqn:OmegaFunLos} implicitly indicates that $\Omega(y^{-\frac{\alpha}{2}})$ is significantly dependent upon the mean of $H$, especially when $y\ell^{\frac{2}{\alpha}}$ is small. Namely, in general $\Omega(y^{-\frac{\alpha}{2}})$ in~\eqref{Eqn:OmegaFunLos} is insensitive to the different distributions of $H$ that have the same mean. This point will be illustrated in Section~\ref{SubSec:SimAPDL}.

\noindent These above observations learned from $R_{\star}$ considerably help us understand some fundamental and intrinsic properties of $\Phi_u$ and they are very useful for the following analyses.

\subsection{Laplace Transforms of the 3D Shot Signal Processes in $\Phi_u$}\label{SubSec:LapTranSSS}

Let the Laplace transform of a non-negative RV $Z$ be defined as $\mathcal{L}_Z(s)\defn \mathbb{E}[\exp(-sZ)]$ for $s>0$. In this subsection, we would like to first study the Laplace transform of the following RV $T_0$ defined as
\begin{align}\label{Eqn:ComSSP}
T_0\defn \sum_{i:U_i\in\Phi_u} W_iL_i\|U_i\|^{-\alpha},
\end{align}
which is referred to as a (complete) 3D (Poisson) shot signal process since it is the sum of all the weighted signal measures in a 3D Poisson field of transmitting points\cite{SBLMCT90,CHLHMH19,CHL19}. Study the Laplace transform of $T_0$ gives rise to some useful results that can be employed to the following coverage analyses of a UAV-enabled cellular network in Section~\ref{Sec:UAVNetworkModel} as the proposed 3D point process $\Phi_u$ is applied to model the locations of UAVs hovering in the sky. Our findings for $\mathcal{L}_{T_0}(s)$ are summarized in the following theorem. 
\begin{theorem}\label{Thm:LapTran1}
Suppose the moment and the Laplace transform of $W_i$ exist for all $i\in\mathbb{N}_+$. (i) If the APIL scenario is considered such that $\Theta_i$ and $X_j$ are independent for all $i, j\in\mathbb{N}_+$, $\mathcal{L}_{T_0}(s)$ can be found as
\begin{align}\label{Eqn:LapTranComSSP1}
\mathcal{L}_{T_0}(s) = \exp\left\{-\pi\lambda s^{\frac{2}{\alpha}}\mathbb{E}\left[W^{\frac{2}{\alpha}}\right]\Gamma\left(1-\frac{2}{\alpha}\right)\omega\right\},
\end{align}
where $\Gamma(z)\defn \int_0^{\infty} y^{z-1}e^{-y}\dif y$ for $z>0$ is the Gamma function. (ii) On the other hand, if the APDL scenario is considered such that $\Theta_k=\tan^{-1}(H_k/\|X_k\|)$ for all $k\in\mathbb{N}_+$, $\mathcal{L}_{T_0}(s)$ can be derived as
\begin{align}\label{Eqn:LapTranComSSP2}
\mathcal{L}_{T_0}(s)=\exp\left(-\pi\lambda \int_0^{\infty}\mathcal{J}_W\left(sz^{-\frac{\alpha}{2}},\tan^{-1}\left(\frac{H}{\sqrt{z}}\right)\right)\dif z\right),
\end{align}
where $\mathcal{J}_W(x,Y)$ for $x,Y>0$ is defined as
\begin{align}
\mathcal{J}_W(x,Y)\defn& \mathbb{E}_Y\big\{ \rho\left(Y\right)\left[1-\mathcal{L}_W\left(x\cos^{\alpha}(Y)\right)\right]\nonumber\\
&+[1-\rho\left(Y\right)] \left[1-\mathcal{L}_W\left(x\ell\cos^{\alpha}(Y)\right)\right]\big\}. \label{Eqn:WfunIndep}
\end{align} 
\end{theorem}
\begin{IEEEproof}
See Appendix~\ref{App:ProofLapTran1}.
\end{IEEEproof} 
\noindent Note that~\eqref{Eqn:LapTranComSSP2} reduces to its closed-form version in~\eqref{Eqn:LapTranComSSP1} once the dependence between the elevation angle and the projection of each point in $\Phi_u$ does not exit. We can infer the distribution of $T_0$ from Theorem~\ref{Thm:LapTran1}. Let $f_{T_0}(\cdot)$ denote the probability density function (PDF) of $T_0$ and it can be obtained by finding the inverse Laplace transform of $T_0$. Namely, by letting $\mathcal{L}^{-1}\{g(s)\}(z)$ denote the inverse Laplace transform of function $g(s)$, we can express the PDF of $T_0$ for the result in~\eqref{Eqn:LapTranComSSP1} as
\begin{align}\label{Eqn:pdfComSSP1}
f_{T_0}(z) = \mathcal{L}^{-1}\left\{\exp\left[-\pi\lambda s^{\frac{2}{\alpha}}\mathbb{E}\left[W^{\frac{2}{\alpha}}\right]\Gamma\left(1-\frac{2}{\alpha}\right)\omega \right]\right\}(z),
\end{align}  
which cannot be further found in closed  form if $\alpha\neq 4$, yet it can be evaluated by numerical techniques. For $\alpha=4$, the  closed-form expression of $f_{T_0}(z)$ can be found as~\cite{CHLHMH19}\cite{MAIA12}
\begin{align}
f_{T_0}(z) = \frac{\pi\mathbb{E}[\sqrt{W}]\lambda\omega}{2z^{\frac{3}{2}}}\exp\left(-\frac{\pi^3(\mathbb{E}[\sqrt{W}]\lambda\omega)^2}{4 z}\right),
\end{align}
which is essentially the PDF of a L\'{e}vy RV with location parameter zero and scale parameter $\pi^3(\mathbb{E}[\sqrt{W}]\lambda\omega)^2/2$. Similarly, the PDF of $T_0$ for the result in~\eqref{Eqn:LapTranComSSP2} can also be expressed as
\begin{align}
f_{T_0}(z) = & \mathcal{L}^{-1}\bigg\{\exp\bigg(-\pi\lambda \int_0^{\infty} \mathcal{J}_W\left(sz^{-\frac{\alpha}{2}},\tan^{-1}\left(\frac{H}{\sqrt{z}}\right)\right)\nonumber\\ 
&\dif z \bigg) \bigg\}(z),
\end{align}
which does not have a closed-form solution and can only be evaluated by numerical methods. 
 
Next, let us define the $K$th-truncated shot signal process in $\Phi_u$ as follows
\begin{align}\label{Eqn:TrucKthSSS}
T_K\defn  \sum_{k=K+1}^{\infty}W_kL_k\|U_k\|^{-\alpha},
\end{align}
where $U_k$ denotes the $k$th nearest point in $\Phi_u$ to the origin, $W_k$ and $L_k$ are  non-negative RVs associating with $U_k$ as already defined in~\eqref{Eqn:Weight3DDis}. Since $T_K$ in~\eqref{Eqn:TrucKthSSS} does not contain the weighted signals emitted from the $K$ points in set $\Phi_u$, it is called the $K$th-truncated shot signal process in $\Phi_u$ and it converges to $T_0$ as $K$ goes to zero. The Laplace transforms of $T_K$ in two different scenarios are found in the following theorem.
\begin{theorem}\label{Thm:LapTran2}
Suppose the moment and the Laplace transform of $W_k$ for all $k\in\mathbb{N}_+$ exist. If the APIL scenario is considered such that $\Theta_i$ is independent of $\|X_j\|$ for all $i,j\in\mathbb{N}_+$, the Laplace transform of $T_K$ defined in~\eqref{Eqn:TrucKthSSS} for $K>0$ can be derived as
\begin{align}\label{Eqn:LapTranImSSP1}
\mathcal{L}_{T_K}(s) =& \mathbb{E}_{D_K}\bigg\{\exp\bigg(-\pi\lambda  D_K\mathbb{E}_{\Theta}\bigg[[1-\rho(\Theta)]\times \nonumber\\
& \mathcal{I}_W\left(s\ell\cos^{\alpha}(\Theta)D_K^{-\frac{\alpha}{2}},\frac{2}{\alpha}\right)+\rho(\Theta)\times\nonumber\\
&\mathcal{I}_W\left(s\cos^{\alpha}(\Theta)D_K^{-\frac{\alpha}{2}},\frac{2}{\alpha}\right)+\bigg] \bigg)\bigg\},  
\end{align}
where $D_K\sim\text{Gamma}(K,\pi\lambda)$ is a Gamma RV with shape parameter $K$ and rate parameter $\pi\lambda$, and $\mathcal{I}_W(u,v)$ for $u,v>0$ is defined as
\begin{align}
\mathcal{I}_W(u,v) \defn & u^v \bigg\{\Gamma(1-v)\mathbb{E}\left[W^v\right]\nonumber\\ &-\int_0^{u^{-v}}\left[1-\mathcal{L}_W\left(x^{-\frac{1}{v}}\right)\right] \dif x\bigg\}.\label{Eqn:Ifun}
\end{align}
On the contrary, if the APDL scenario is considered such that $\Theta_i=\tan^{-1}(H_i/\|X_i\|)$,  $\mathcal{L}_{T_K}(s)$ is found as
\begin{align}\label{Eqn:LapTranImSSP2}
\mathcal{L}_{T_K}(s) =  \mathbb{E}_{D_K}\bigg\{ & \exp\bigg[-\pi\lambda  \int_{D_K}^{\infty}\mathcal{J}_W\left(\dfrac{s}{z^{\frac{\alpha}{2}}},\tan^{-1}\left(\frac{H}{\sqrt{z}}\right) \right)\nonumber\\
&\dif z\bigg]\bigg\}.
\end{align}
\end{theorem}
\begin{IEEEproof}
See Appendix~\ref{App:ProofLapTran2}.
\end{IEEEproof}
Although the result in~\eqref{Eqn:LapTranImSSP2} is somewhat complicated due to considering the dependence between the elevation angle and the project of each point in $\Phi_u$, it reduces to the result in~\eqref{Eqn:LapTranImSSP1} once  the dependence no longer exists. In general, \eqref{Eqn:LapTranImSSP1} cannot be further expressed as a closed-form outcome, yet it simply reduces to the following expression for a special case of $s=\zeta \sec^{\alpha}(\Theta)D_K^{\frac{\alpha}{2}}$ for any constant $\zeta>0$:
\begin{align}\label{Eqn:LapTransTurnCloseForm}
\mathcal{L}_{T_K}\left(\zeta\sec^{\alpha}(\Theta) D_K^{\frac{\alpha}{2}}\right)=& \bigg[1+\mathbb{E}\left[\rho(\Theta)\right]\mathcal{I}_W\left(\zeta,\frac{2}{\alpha}\right)+\nonumber\\
&(1-\mathbb{E}\left[\rho(\Theta)\right])\mathcal{I}_W\left(\zeta\ell,\frac{2}{\alpha}\right)\bigg]^{-K},
\end{align}
and we will find this result quite useful for the analyses in the following sections. In general, the PDF of $T_K$ cannot be tractably derived by finding the inverse Laplace transforms of \eqref{Eqn:LapTranImSSP1} and \eqref{Eqn:LapTranImSSP2} thanks to their complicated forms. Nevertheless, we will see that Theorem~\ref{Thm:LapTran2} plays a pivotal role in the following coverage analyses of a UAV-enabled cellular network. 

\section{Modeling and Analysis of a UAV-Enabled Cellular Network Using $\Phi_u$}\label{Sec:UAVNetworkModel}

In this section, we employ the proposed 3D point process $\Phi_u$ in~\eqref{Eqn:3DPP} to model the random locations of UAVs in a cellular network, as shown in Fig.~\ref{Fig:SystemModel}. The salient feature of using $\Phi_u$ to model the 3D locations of the UAVs, as we will see, is not only  to generally characterize the distribution of the UAVs hovering in the sky but also to properly and tractably analyze the performances of a UAV-enabled cellular network. Our focus in this section is on the study of the coverage performance of a UAV-enabled cellular network in which a tier of UAVs are deployed in the sky that serve as \textit{aerial} base stations in the network and the locations of the UAVs are modeled by $\Phi_u$, that is, $U_i$ in $\Phi_u$ denotes UAV $i$ and its location in the network. Note that in this paper our focus is to study how to generally deploy a large-scale UAV-enabled cellular network and analyze its performances of a snapshot in time so that in general the proposed $\Phi_u$ cannot characterize the continuous-time mobility impacts of the UAVs on the network performances. Nonetheless, $\Phi_u$ still works for modeling the positions of all mobile UAVs whose trajectories are quiet different at any particular time point in that they can be properly assumed to be independent and thereby well approximated by $\Phi_u$.

Suppose there is a typical user located at the origin and each user in the UAV-enabled cellular network associates with a UAV that provides it with the (averaged) strongest received signal power. Namely, the UAV associated with the typical user is given by
\begin{align}\label{Eqn:AssoUAV}
 U_{\star}&\defn\arg\max_{i:U_i\in\Phi_u} \mathbb{E}\left[PL_iG_i\|U_i\|^{-\alpha}|U_i\right]\nonumber\\ &=\arg\max_{i:U_i\in\Phi_u} \frac{P\mathbb{E}[G]L_i}{\|U_i\|^{\alpha}}\nonumber\\
 & =\arg\max_{i:U_i\in\Phi_u} \frac{L_i}{\|U_i\|^{\alpha}},
\end{align} 
where $P$ is the transmit power of each UAV, $G_i\sim\exp(1)$ denotes the fading channel gain between the typical user\footnote{The fading channel gain $G_i$ is assumed to be an exponential RV with unit mean and not affected by the $N$ transmit antennas of a UAV because each UAV broadcasts its user association signaling during the phase of user association and thereby it cannot do downlink transmit beamforming to any specific user.} and $U_i$, $\alpha>2$ denotes the path-loss exponent in this context, and $L_i$, as already defined in~\eqref{Eqn:Weight3DDis}, is used to characterize the LoS and NLoS channel effects in the channel between $U_i$ and the typical user. The second equality in~\eqref{Eqn:AssoUAV} is due to considering the independence between $G_i$ and $U_i$ as well as conditioning on $U_i$, and the third equality is owing to  removing constants $P$ and $\mathbb{E}[G]$ does not affect the result of finding $U_{\star}$.

\subsection{The SINR Model}
Let $I_0$ be the aggregated interference power received by the typical user that does not include the signal power from $U_{\star}$ so that it can be written as
\begin{align}\label{Eqn:Interference}
I_0 \defn \sum_{i:U_i\in\Phi_u\setminus U_{\star}} P G_i L_i\|U_i\|^{-\alpha}.
\end{align}
All $G_i$'s are assumed to be i.i.d. and they are independent of all $L_i$'s and $U_i$'s. Note that each UAV is associated with at least one user so that the ``void'' UAV phenomenon is not modeled in $I_0$~\cite{CHLLCW1502}\cite{CHLLCW16}. In addition, each UAV allocates different resource blocks (RBs) to different users associating with it, i.e., no users associating with the same UAV can share the same RB. 

 Each UAV is assumed to be equipped with $N$ antennas whereas each user is equipped with a single antenna. According to~\eqref{Eqn:AssoUAV} and~\eqref{Eqn:Interference}, if each UAV is able to perform transmit beamforming to its user, the signal-to-interference plus noise power ratio (SINR) of the typical user can be defined as 
\begin{align}\label{Eqn:DefSINR}
\gamma_0 \defn \frac{PG_{\star}L_{\star}\|U_{\star}\|^{-\alpha}}{I_0+\sigma_0},
\end{align}
where $G_{\star}\sim\text{Gamma}(N,1)$ is the fading channel gain from $U_{\star}$ to the typical user\footnote{The fading channel gain $G_{\star}$ is assumed to be a Gamma RV with shape parameter $N$ and rate parameter $1$ (i.e., $G_{\star}\sim\text{Gamma}(N,1)$) because UAV $U_{\star}$ is serving the typical user so that it knows the channel state information (CSI) of the typical user and is thus able to do downlink transmit beamforming to the typical user. Hence, the mean of $G_{\star}$ is $\mathbb{E}[G_{\star}]=N$. All the fading channel gains in $I_0$ can be shown to be i.i.d. exponential RVs with unit mean (i.e., $G_i\sim\exp(1)$) since all the interfering UAVs do not know the CSI from them to the typical user and are thus unable to do transmit beamforming to the typical user. For the detailed explanation about how to derive $G_{\star}\sim\text{Gamma}(N,1)$ and $G_i\sim\exp(1)$, please refer to Appendix A in~\cite{HSDMKJA13} or Section II-D in~\cite{PXCHLJA13}.}, $L_{\star}\in\{1,\ell\}$ has the same distribution as $L_i$, and $\sigma_0$ denotes the thermal noise power from the environment. The downlink coverage (probability) of a user in the network can thus be defined as
\begin{align}\label{Eqn:KjointCovProb}
p_{cov}\defn \mathbb{P}\left[\gamma_0 \geq \beta\right]= \mathbb{P}\left[\frac{ PG_{\star}L_{\star}\|U_{\star}\|^{-\alpha}}{I_0+\sigma_0}\geq \beta \right],
\end{align}
where $\beta>0$ is the SINR threshold for successful decoding. In the following, we will analyze $p_{cov}$ by considering whether the elevation angle and the projection of each UAV are independent or not. In the following two sections, we will employ the model of the UAV-enabled cellular network proposed in this section to analyze the coverage performances of the network in the APIL and APDL scenarios.
 
 \subsection{Downlink Coverage Analysis: The APIL Scenario}\label{SubSec:PerAnaAPIL}
 
In this subsection, we would like to study the downlink coverage $p_{cov}$ in~\eqref{Eqn:KjointCovProb} by considering the APIL scenario, i.e., elevation angle $\Theta_i$ and projection $X_i$ of UAV $U_i$ are independent for all $i\in\mathbb{N}_+$. The following proposition, which is developed by employing Theorem~\ref{Thm:LapTran2} to a first-truncated shot signal process in the 3D point process $\widetilde{\Phi}_u$ defined in~\eqref{Eqn:Equ3DPointProcess}, specifies the analytical result of $p_{cov}$ in this scenario.
\begin{proposition}\label{Prop:CovProbIndep}
If the APIL scenario is considered, the downlink coverage defined in~\eqref{Eqn:KjointCovProb} can be found as
\begin{align}\label{Eqn:CovProbIndep}
p_{cov} =& \frac{\dif^{N-1}}{\dif t^{N-1}}\mathbb{E}\bigg[\frac{t^{N-1}}{(N-1)!}\exp\bigg(-\frac{\sigma_0 D_{\star}^{\frac{\alpha}{2}}}{t P}\nonumber\\
&-\pi\lambda\omega D_{\star}\mathcal{I}_G\left(\frac{1}{t},\frac{2}{\alpha}\right)\bigg)\bigg]\bigg|_{t=\frac{1}{\beta}}, 
\end{align} 
where $D_{\star}\sim\exp(\pi\lambda\omega)$ and function $\mathcal{I}_G(u,v)$ is defined as
\begin{align}
\mathcal{I}_G\left(u,v\right)\defn u^{v}\left(\frac{\pi v}{\sin(\pi v)}-\int_0^{u^{-v}}\frac{\dif r}{1+r^{\frac{1}{v}}}\right). \label{Eqn:IfunNoW}
\end{align}
\end{proposition}
\begin{IEEEproof}
See Appendix~\ref{App:ProofCovProbIndep}.
\end{IEEEproof}
\noindent We adopt an exponential RV $D_{\star}$ with mean $1/\pi\lambda\omega$ in~\eqref{Eqn:CovProbIndep} to make $p_{cov}$ show in a neat form so as to clearly see how $p_{cov}$ is impacted by $D_{\star}$ and other network parameters. The physical meaning of $D_{\star}$ is the square of the shortest distance between the typical user and set $\widetilde{\Phi}_u$, i.e., $D_{\star}\defn\|\widetilde{U}_{\star}\|^2\stackrel{d}{=}L^{-\frac{2}{\alpha}}_{\star}\|U_{\star}\|^2$ where $\widetilde{U}_{\star}$ is the nearest point in $\widetilde{\Phi}_u$ to the typical user and $\stackrel{d}{=}$ stands for the equivalence in distribution. In other words, $p_{cov}$ is highly dependable upon the distribution of elevation angle $\Theta$ and $\ell$ for a given density $\lambda$ because the distribution of $D_{\star}$ is parameterized with $\lambda\omega$. To make this point much clear, we use Jensen's inequality to find a lower bound on $p_{cov}$ in~~\eqref{Eqn:CovProbIndep} as
\begin{align}\label{Eqn:CovProbIndepIneq1}
p_{cov}\geq & \frac{1}{(N-1)!}\frac{\dif^{N-1}}{\dif t^{N-1}}\bigg\{t^{N-1}\exp\bigg[-\frac{\sigma_0\Gamma\left(1+\frac{\alpha}{2}\right)}{t P(\pi\lambda\omega)^{\frac{\alpha}{2}}}\nonumber\\
&-\mathcal{I}_G\left(\frac{1}{t},\frac{2}{\alpha}\right)\bigg]\bigg\}\bigg|_{t=\frac{1}{\beta}},
\end{align}
which reduces to the following neat inequality for $N=1$:
\begin{align}\label{Eqn:CovProbIndepIneq2}
p_{cov}\geq \exp\left[-\frac{\beta\sigma_0\Gamma\left(1+\frac{\alpha}{2}\right)}{ P(\pi\lambda\omega)^{\frac{\alpha}{2}}}-\mathcal{I}_G\left(\beta,\frac{2}{\alpha}\right)\right].
\end{align}
The inequalities in~\eqref{Eqn:CovProbIndepIneq1} and \eqref{Eqn:CovProbIndepIneq2}  apparently show that increasing $\lambda\omega$ improves $p_{cov}$. This is because users are able to associate with a nearer UAV and receive stronger power from the UAV when deploying UAVs more densely even though more interference is generated as well. Also, $p_{cov}$ improves whenever  $\lambda\omega$ can be maximized by optimizing the distribution of $\Theta$. We will demonstrate some numerical results in Section~\ref{Sec:Simulation} to show how $p_{cov}$ varies with different distribution cases of $\Theta$. However, when the network is interference-limited (i.e., $\sigma_0=0$), $p_{cov}$ in~\eqref{Eqn:CovProbIndep} significantly reduces to the following expression 
\begin{align}\label{Eqn:CovProbIntLimit}
p_{cov} &= \frac{1}{(N-1)!}\frac{\dif^{N-1}}{\dif t^{N-1}}\left(\frac{t^{N-1}}{1+\mathcal{I}_G\left(\frac{1}{t},\frac{2}{\alpha}\right)}\right)\bigg|_{t=\frac{1}{\beta}}\nonumber\\
&\stackrel{(N=1)}{=} \frac{1}{1+\mathcal{I}_G(\beta,\frac{2}{\alpha})},
\end{align}
and further reduces to a closed-form result as $N=1$, which is not impacted by $\lambda\omega$.  Thus, we can draw a conclusion that the downlink coverage tends to be more sensitive to the distribution of the elevation angle and the density of the projections of the UAVs as the network tends to be more ``noise-limited" (i.e., noise power dominates the SINR performance). Moreover, as $N$ goes to infinity, $p_{cov}$ in~\eqref{Eqn:CovProbIndep} increases up to the following limit
\begin{align}
p_{cov,\infty} \defn& \lim_{N\rightarrow\infty} p_{cov}\nonumber\\
 = & \mathcal{L}^{-1}\bigg\{ \mathbb{E}\bigg[\frac{1}{s}\exp\bigg(- \frac{s\sigma_0D_{\star}^{\frac{\alpha}{2}}}{P}-\pi\lambda \omega D_{\star}\nonumber\\
 & \times\mathcal{I}_G\left(s,\frac{2}{\alpha}\right)\bigg) \bigg]\bigg\}\left(\frac{1}{\beta}\right), \label{Eqn:ProofCovProbIndeInfN}
\end{align}
which is the upper limit of the downlink coverage for a user associating with a single UAV with a massive antenna array.

An effective method to significantly improve the coverage of users is to make users associate with multiple UAVs so that the UAVs can do coordinated multi-point (CoMP) joint transmission. The upper limit of the downlink coverage of a user associating with multiple UAVs can be achieved when all the UAVs are coordinated to jointly transmit to the user at the same time, which is referred as to the \textit{cell-free downlink coverage}. Since perfectly coordinating and synchronizing all the UAVs in a large-scale network to do \textit{coherent} transmission is hardly possible in practice, \textit{non-coherent joint transmission} is a feasible way for all the UAVs to jointly achieve the cell-free downlink coverage in that it has lower implementation complexity and does not require large backhaul capacity if compared with its coherent counterpart\footnote{Studying the fundamental limit of the downlink coverage of a UAV-enabled cellular network is the main purpose in this paper. Accordingly, in the following analysis we merely analyze how much cell-free downlink coverage can be achieved when all the UAV adopt non-coherent joint transmission to serve one user. The cell-free downlink coverage problem of multiple users served by all the UAVs and its related practical issues are beyond the scope of this paper and they are left for our future study.}. When all the UAVs perform non-coherent CoMP joint transmission to a user, the cell-free downlink coverage of the user can be defined as~\cite{DLHSBCEH12,RTSSJGA14}
\begin{align}\label{Eqn:DefnCellFreeCovProb}
p^{cf}_{cov} \defn \mathbb{P}\left[\frac{P\sum_{i:U_i\in\Phi_u}G_iL_i\|U_i\|^{-\alpha}}{\sigma_0} \geq \beta \right],
\end{align}
where $G_i\sim\text{Gamma}(N,1)$ for all $i\in\mathbb{N}_+$ since all the UAVs can do transmit beamforming to the user. The explicit result of $p^{cf}_{cov}$ can be found by using Theorem~\ref{Thm:LapTran1} and it is shown in the following proposition.
\begin{proposition}\label{Prop:CellFreeCovProbAPIL}
If all the UAVs are deployed based on the APIL scenario and coordinated to do non-coherence joint transmission, the cell-free downlink coverage defined in~\eqref{Eqn:DefnCellFreeCovProb} is derived as
\begin{align}\label{Eqn:CellFreeCov1}
p^{cf}_{cov}= 1-\mathcal{L}^{-1}\bigg\{ &\frac{1}{s}\exp\bigg[-\frac{\pi\lambda s^{\frac{2}{\alpha}}\omega}{(N-1)!}\Gamma\left(N+\frac{2}{\alpha}\right)\nonumber\\
&\Gamma\left(1-\frac{2}{\alpha}\right)\bigg]\bigg\}\left(\frac{\beta\sigma_0}{P}\right),
\end{align}
which reduces to the following closed-form result for $\alpha=4$:
\begin{align}\label{Eqn:CellFreeCov2}
p^{cf}_{cov} = \mathrm{erf}\left(\frac{\pi^{\frac{3}{2}}\lambda\omega}{2(N-1)!}\sqrt{\frac{P}{\beta\sigma_0}}\Gamma\left(N+\frac{1}{2}\right)\right),
\end{align}
where $\mathrm{erf}(z)\defn \frac{2}{\sqrt{\pi}}\int_0^z e^{-t^2}\dif t$ is the error function for $z>0$.
\end{proposition}
\begin{IEEEproof}
See Appendix~\ref{App:ProofCellFreeCoverage}.
\end{IEEEproof}
The cell-free downlink coverage in \eqref{Eqn:DefnCellFreeCovProb} can be interpreted as the maximum downlink coverage jointly achieved by all the UAVs with $N$ antennas. When $N$ goes to infinity, \eqref{Eqn:CellFreeCov1} approaches its upper limit given by
\begin{align}\label{Eqn:CellFreeCovInfN1}
 p^{cf}_{cov,\infty} &\defn\lim_{N\rightarrow\infty} p^{cf}_{cov}\nonumber\\
 &= 1-\mathcal{L}^{-1}\left\{\frac{1}{s}\exp\left[-\pi\lambda s^{\frac{2}{\alpha}}\omega\Gamma\left(1-\frac{2}{\alpha}\right)\right]\right\}\left(\frac{\beta\sigma_0}{P}\right),
\end{align} 
which reduces to the following closed-form results for $\alpha=4$:
\begin{align}\label{Eqn:CellFreeCovInfN2}
 p^{cf}_{cov,\infty}=\lim_{N\rightarrow\infty} p^{cf}_{cov} = \mathrm{erf}\left(\frac{\pi^{\frac{3}{2}}\lambda\omega}{2}\sqrt{\frac{P}{\beta\sigma_0}}\right).
\end{align}
The cell-free downlink coverage in~\eqref{Eqn:CellFreeCovInfN1} and its closed-form special case in~\eqref{Eqn:CellFreeCovInfN2} are the fundamental limit of the downlink coverage achieved by all the UAVs that are equipped with a massive antenna array and perform non-coherent joint transmission and this fundamental limit is referred to as \textit{the cell-free massive MIMO coverage} of a UAV-enabled cellular network in the APIL scenario. To the best of our knowledge, they are firstly derived in this paper. Note that  $p^{cf}_{cov}$ is dominated by $\lambda\omega$ and $P$ and increasing $\lambda\omega$ is more efficient to improve it than increasing $P$, and thereupon it is also significantly affected by the distribution of the elevation angle of the UAVs. Hence, optimizing the distribution of the elevation angle of each UAV may also considerably improve $p^{cf}_{cov}$, which will be numerically demonstrated in Section~\ref{Sec:Simulation}.

\subsection{Downlink Coverage Analysis of a UAV-Enabled Network: The APDL Scenario}\label{SubSec:PerAnaAPDL}

In the subsection, we turn our focus to the downlink coverage in the APDL scenario where the elevation angle and the projection of each UAV are dependent. For $\Theta_i=\tan^{-1}(H_i/\|X_i\|)$ for each UAV $U_i$, the following theorem summarizes the explicit expression of the downlink coverage in this scenario. 
\begin{proposition}\label{Prop:CovProbDep}
	If the APDL scenario is considered, the downlink coverage defined in~\eqref{Eqn:KjointCovProb} can be found as
	\begin{align}\label{Eqn:CovProbAPDL}
	p_{cov}= &\frac{\dif^{N-1}}{\dif t^{N-1}}\mathbb{E}_{\widetilde{D}_{\star}}\bigg\{\frac{t^{N-1}}{(N-1)!}\exp\bigg[-\frac{\sigma_0 \widetilde{D}^{\frac{\alpha}{2}}_{\star}}{P t}-\pi\lambda\nonumber\\
	&\times\widetilde{D}_{\star} \widetilde{\mathcal{I}}_{G}\left(\frac{1}{t},\frac{2}{\alpha},\widetilde{D}_{\star}\right)\bigg]\bigg\}\bigg|_{t=\frac{1}{\beta}},
	\end{align}
	where $\widetilde{D}_{\star}$ is a non-negative RV with the following PDF
	\begin{align}\label{Eqn:PDFTildeDstar}
	f_{\widetilde{D}_{\star}} (y) = \pi\lambda \Omega'\left(y^{-\frac{\alpha}{2}}\right)e^{-\pi\lambda\Omega\left(y^{-\frac{\alpha}{2}}\right)}
	\end{align}
	with $\Omega\left(y^{-\frac{\alpha}{2}}\right)$ is defined in~\eqref{Eqn:OmegaFunLos}  for the APDL scenario, $\Omega'\left(y^{-\frac{\alpha}{2}}\right)\defn\frac{\dif\Omega\left(y^{-\frac{\alpha}{2}}\right)}{\dif y}$, and $\widetilde{\mathcal{I}}_G(u,v,r)$ for $u,v,r>0$ is defined as
	\begin{align}
	\widetilde{\mathcal{I}}_G(u,v,r) \defn \int_1^{\infty} \Omega'\left(\frac{r}{y^v}\right)\left(\frac{u}{y^{\frac{1}{v}}+u}\right) \dif y. \label{Eqn:IfunforAPDL}
	\end{align}
\end{proposition}
\begin{IEEEproof}
	See Appendix \ref{App:ProofCovProbDep}.
\end{IEEEproof}
\noindent Note that the physical meaning of $\widetilde{D}_{\star}$ is the square of the short distance from the typical user to set $\widetilde{\Phi}_u$ in~\eqref{Eqn:Equ3DPointProcess} in the APDL scenario and $p_{cov}$ in~\eqref{Eqn:CovProbAPDL} becomes $p_{cov}$ in~\eqref{Eqn:CovProbIndep} as $\widetilde{D}_{\star}$ in~\eqref{Eqn:CovProbAPDL} reduces to $D_{\star}$ in~\eqref{Eqn:CovProbIndep} (i.e., $\Omega(z)$ reduces to $\omega z$). Applying the Jensen inequality on $\widetilde{D}_{\star}$ in~\eqref{Eqn:CovProbAPDL} gives rise to the following lower bound on $p_{cov}$ in~\eqref{Eqn:CovProbAPDL}:
\begin{align}
p_{cov}\geq & \frac{\dif^{N-1}}{\dif t^{N-1}} \bigg\{\frac{t^{N-1}}{(N-1)!}\exp\bigg[-\frac{\sigma_0 \mathbb{E}\left[\widetilde{D}^{\frac{\alpha}{2}}_{\star}\right]}{P t}-\pi\lambda\nonumber\\
& \times\mathbb{E}[\widetilde{D}_{\star}] \widetilde{\mathcal{I}}_{G}\left(\frac{1}{t},\frac{2}{\alpha}\right)\bigg]\bigg\}\bigg|_{t=\frac{1}{\beta}}
\end{align}
and for the interference-limited situation it reduces to
\begin{align}
p_{cov}\geq \frac{\dif^{N-1}}{\dif t^{N-1}} \bigg\{ & \frac{t^{N-1}}{(N-1)!}\exp\bigg[-\pi\lambda \mathbb{E}\left[\widetilde{D}_{\star}\right]\nonumber\\ &\times\widetilde{\mathcal{I}}_{G}\left(\frac{1}{t},\frac{2}{\alpha}\right)\bigg]\bigg\}\bigg|_{t=\frac{1}{\beta}}.
\end{align}
According to the PDF of $\widetilde{D}_{\star}$ in~\eqref{Eqn:PDFTildeDstar}, we know $\mathbb{E}[\Omega(\widetilde{D}^{-\frac{\alpha}{2}}_{\star})]=\frac{1}{\pi\lambda}$ so that  $\Omega(\mathbb{E}[\widetilde{D}^{-\frac{\alpha}{2}}_{\star}])\leq \frac{1}{\pi\lambda}$ or $\Omega(\mathbb{E}[\widetilde{D}^{-\frac{\alpha}{2}}_{\star}])\geq \frac{1}{\pi\lambda}$ depending on the convexity of $\Omega(\cdot)$.  In other words, $\lambda\mathbb{E}[\widetilde{D}^{\frac{\alpha}{2}}_{\star}]$ is still pertaining to $\lambda$ and $\Omega(y^{-\frac{\alpha}{2}})$ so that the UAV density impacts the downlink coverage no matter whether or not the network is interference-limited, which is quite different from $p_{cov}$ in~\eqref{Eqn:CovProbIndep}. Therefore, properly deploying UAVs depending on the distribution of the altitude $H$ of the UAVs is able to reduce $\lambda\Omega(y^{-\frac{\alpha}{2}})$ so as to improve the downlink coverage in the APDL scenario. For example, if each user associates with its nearest UAV with altitude $H\stackrel{d}{=}D_{\star}$ ( i.e., $L_i=1$ and $H\sim\exp(\pi\lambda)$ for all $i\in\mathbb{N}_+$), $\Omega(y^{-\frac{\alpha}{2}})=F_H(y)=1-\exp(-\pi\lambda y)$ and $\Omega'(y^{-\frac{\alpha}{2}})=f_H(y) =\pi\lambda \exp(-\pi\lambda y)$ based on the discussions in Section~\ref{SubSec:Dis-RelDis}. Thus, $\mathbb{E}\left[\widetilde{D}^{\frac{\alpha}{2}}_{\star}\right]$ and $\lambda\mathbb{E}[\widetilde{D}_{\star}]$ decrease as $\lambda$ increases such that $p_{cov}$ always improves as more UAVs are deployed in this example. In addition, $p_{cov}$ in~\eqref{Eqn:CovProbAPDL} increases to its upper limit as the number of antennas equipped at each UAV goes to infinity, which can be shown by using the technique of inverse Laplace transform as used in~\eqref{Eqn:ProofCovProbIndeInfN}.

Next, we would like to study how much the downlink coverage can be achieved when all the UAVs can  perform the aforementioned non-coherent joint transmission in the previous subsection, i.e., the cell-free downlink coverage defined in~\eqref{Eqn:DefnCellFreeCovProb} for the APDL scenario. The following proposition shows its explicit result. 
\begin{proposition}\label{Prop:CellFreeCovProbAPDL}
	If all the UAVs are deployed based on the APDL scenario and coordinated to do non-coherence joint transmission, the cell-free downlink coverage in~\eqref{Eqn:DefnCellFreeCovProb} can be derived as
	\begin{align}\label{Eqn:CellFreeCovProbAPDL}
	p^{cf}_{cov} = 1- &\mathcal{L}^{-1}\bigg\{\frac{1}{s}\exp\bigg[-\pi\lambda \int_0^{\infty}\mathcal{J}_G\bigg(sz^{-\frac{\alpha}{2}},\nonumber\\
	&\tan^{-1}\left(\frac{H}{\sqrt{z}}\right)\bigg)\dif z\bigg]\bigg\}\left(\frac{\beta\sigma_0}{P}\right),
	\end{align}
	where $\mathcal{J}_G(x,Y)$ is
	\begin{align}
	\mathcal{J}_G(x,Y)=& 1- \mathbb{E}_Y\bigg\{ \rho\left(Y\right)\left(1+\frac{x\cos^{\alpha}(Y)}{N}\right)^{-N}\nonumber\\
	&+[1-\rho(Y)]\left(1+\frac{x\cos^{\alpha}(Y)\ell}{N}\right)^{-N}\bigg\}.\label{Eqn:JfunAPDL1}
	\end{align}
	Also, as $N\rightarrow\infty$, we have
	\begin{align}
	\lim_{N\rightarrow\infty}\mathcal{J}_G(x,Y)=1- &\mathbb{E}_Y\bigg\{\rho(Y)e^{-x\cos^{\alpha}(Y)}+[1-\rho(Y)]\nonumber\\
	&\times e^{-x\ell\cos^{\alpha}(Y)}\bigg\}. \label{Eqn:JfunAPDL2}
	\end{align}
\end{proposition}
\begin{IEEEproof}
	The proof is omitted since it is similar to the proof of Proposition~\ref{Prop:CellFreeCovProbAPIL}.
\end{IEEEproof}

\begin{table*}[h] 
	\centering
	\caption{Network Parameters for Simulation~\cite{AHKSSL14}}\label{Tab:SimPara}
	\begin{tabular}{|c|c|}
		\hline Transmit Power  (mW)  $P$  & $50$ \\ 
		\hline Density of set $\Phi_x$ (points (UAVs)/m$^2$) $\lambda_x$  & $1.0\times 10^{-7}\sim 1.0\times 10^{-5}$ (or see figures)     \\ 
		\hline Number of Antennas $N$ & $1$, $4$, $8$, $\infty$ (or see figures) \\ 
		\hline Noise Power (dBm) $\sigma_0$ & $-92.5$ \\ 
		\hline Path-loss Exponent $\alpha$ & $2.75$\\ 
		\hline Parameters $(c_1,c_2)$ in \eqref{Eqn:LoSProb} for Suburban & $(24.5811,39.5971)$ \\
		\hline NLoS Channel Attenuation Factor $\ell$ & $0.25$ \\
		\hline SINR Threshold (dB) $\beta$  & $-10$ (or see figures) \\ 
		\hline
	\end{tabular} 
\end{table*}

\begin{figure*}[t!]
	\centering
	\includegraphics[width=1\textwidth,height=3.0in]{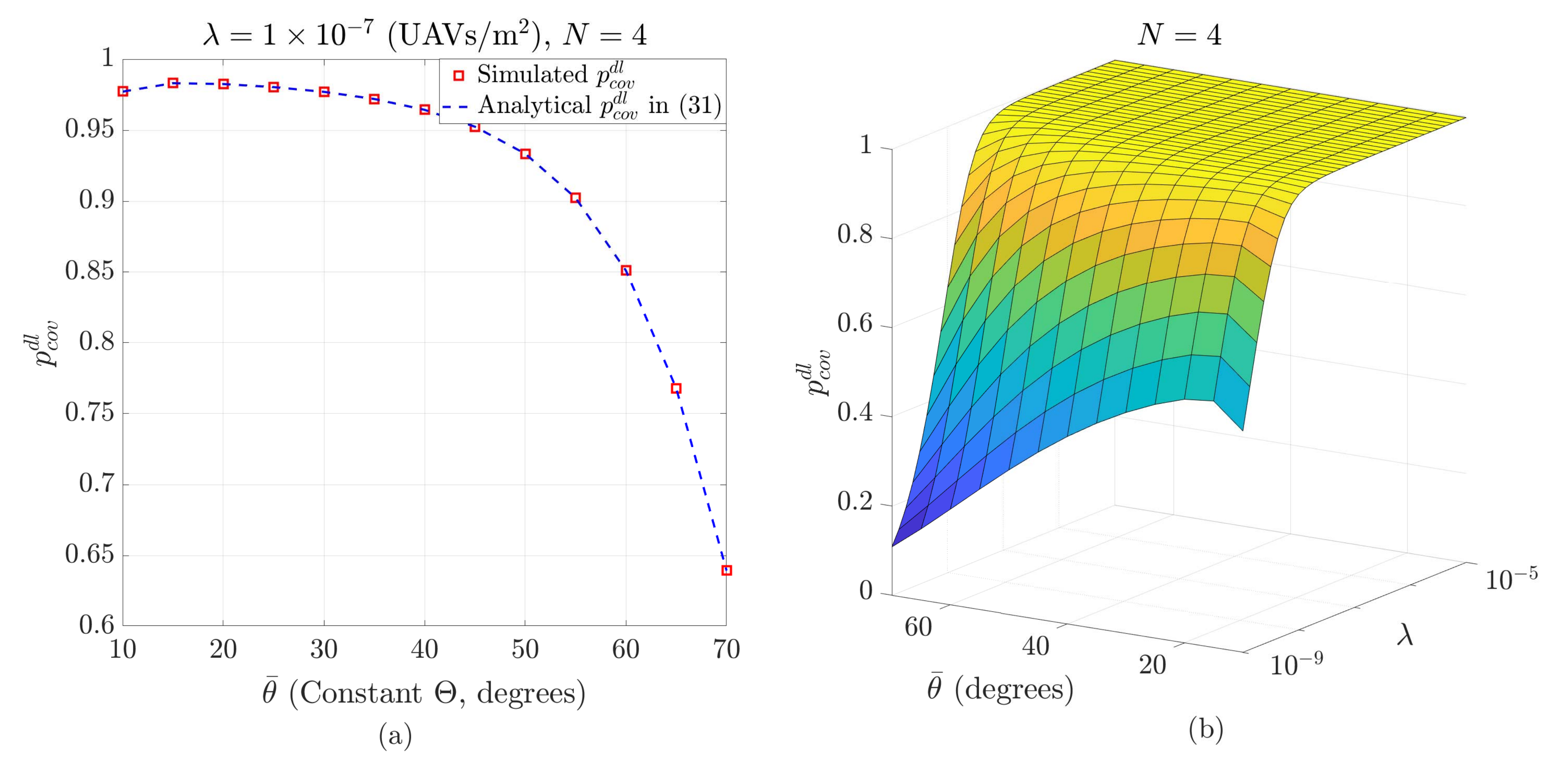}
	\caption{Simulation results of $p^{dl}_{cov}$ for the APIL scenario when $N=4$ and the elevation angle  of each UAV is a constant $\overline{\theta}$ with respect to the origin (i.e., $\tan(\Theta)\sim\text{Gamma}(a,a/\tan(\overline{\theta}))$ as $a\rightarrow\infty$): (a) The 2D simulation results of $p^{dl}_{cov}$ versus elevation angle $\overline{\theta}$ for $\lambda=1\times 10^{-7}$ (UAV /m$^2$), (b) The 3D simulation results of $p^{dl}_{cov}$ versus density $\lambda$ and elevation angel $\overline{\theta}$.}
	\label{Fig:APIL_ConstAng}
\end{figure*}

\noindent Note that \textit{the cell-free massive MIMO coverage} $p^{cf}_{cov,\infty}$ can be readily found by substituting~\eqref{Eqn:JfunAPDL2} into~\eqref{Eqn:CellFreeCovProbAPDL} and it is the fundamental limit of the downlink coverage achieved in a UAV-enabled cellular networks for the APDL scenario. In general, the closed form of $p_{cov}$~\eqref{Eqn:CellFreeCovProbAPDL} cannot be derived, yet it does exist in some special cases. For example, when the altitudes of the UAVs are controlled such that they are proportional to their projection distance (i.e., $H_i=h_0\|X_i\|$ for some $h_0>0$ and all $i\in\mathbb{N}_+$), $\alpha=4$ and $\ell=1$, $p^{cf}_{cov}$ in~\eqref{Eqn:CellFreeCovProbAPDL} reduces to the following closed-form expression
\begin{align}
p^{cf}_{cov} = \mathrm{erf}\left(\frac{\pi \lambda}{2}\sqrt{\frac{P}{\beta\sigma_0}}\int_0^{\infty}\mathcal{J}_G\left(\frac{1}{z^2},\tan^{-1}(h_0)\right)\dif z \right),
\end{align}
where $\mathcal{J}_G(z^{-2},\tan^{-1}(h_0)) = 1-\rho(\tan^{-1}(h_0))(1+\cos^4(\tan^{-1}(h_0))/z^2N)^{-N}$. Also note that all the analytical outcomes in this section are valid as long as $\alpha$ is greater than two, which works for most practical 3D path-loss channel models.  In the following section, we will present some numerical results to verify the above analytical findings of the downlink coverage.

\section{Numerical Results and Discussions}\label{Sec:Simulation}

\begin{figure*}[t!]
	\centering
	\includegraphics[width=1\textwidth,height=3.0in]{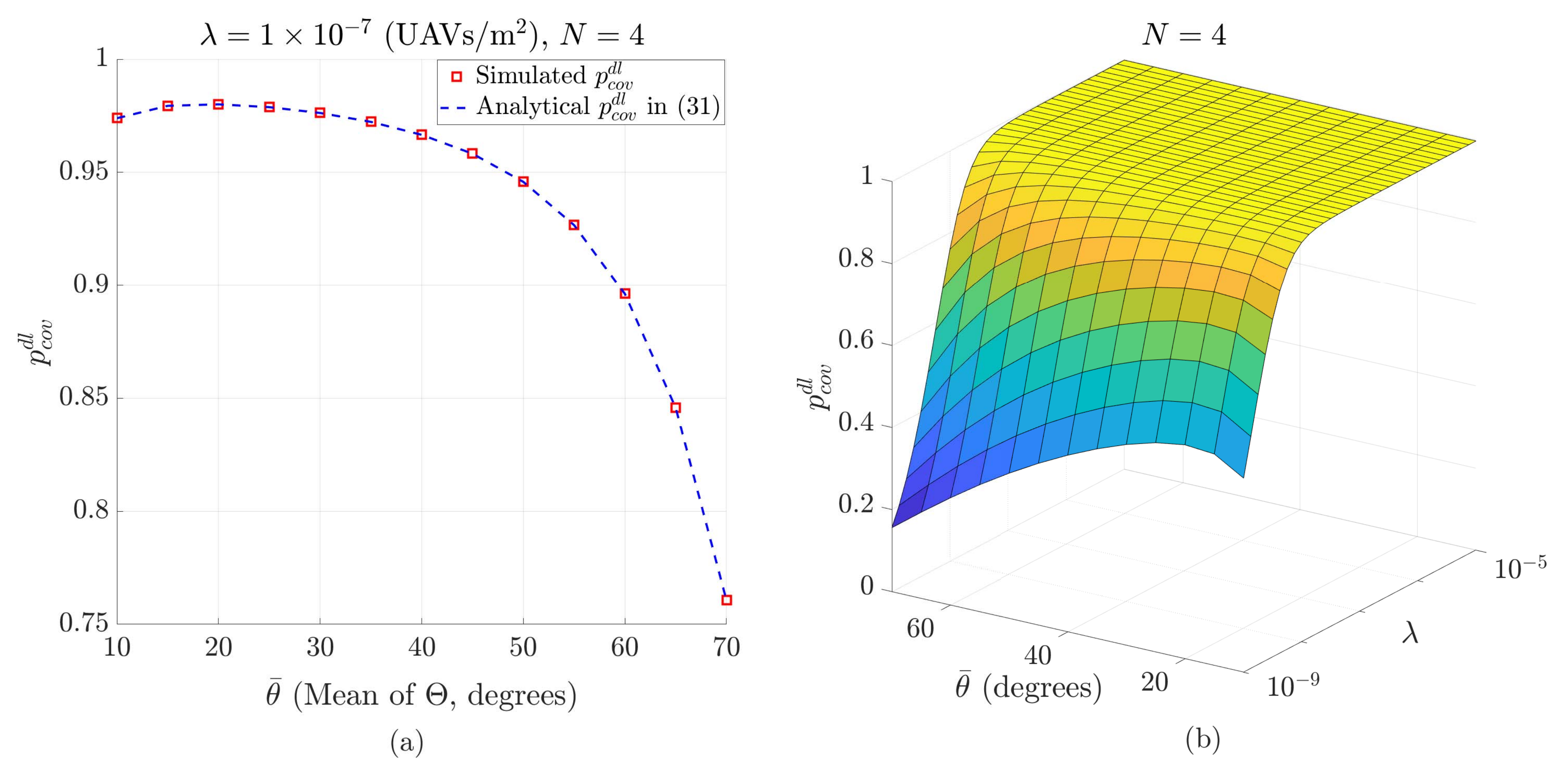}
	\caption{Simulation results of $p^{dl}_{cov}$ for the APIL scenario when $N=4$ and the elevation angle of each UAV is a Gamma RV with shape parameter $a$ and rate parameter $a/\tan(\overline{\theta})$, i.e., $\tan(\Theta)\sim\text{Gamma}(a,a/\tan(\overline{\theta}))$: (a) The 2D simulation results of $p^{dl}_{cov}$ versus elevation angle $\Theta$ for $\lambda=1\times 10^{-7}$ (UAV/m$^2$), (b) The 3D simulation results of $p^{dl}_{cov}$ versus density $\lambda$ and mean of elevation angel $\overline{\theta}$.}
	\label{Fig:APIL_GammaAng}
\end{figure*}

\begin{figure*}[t!]
	\centering
	\includegraphics[width=1\textwidth,height=3.0in]{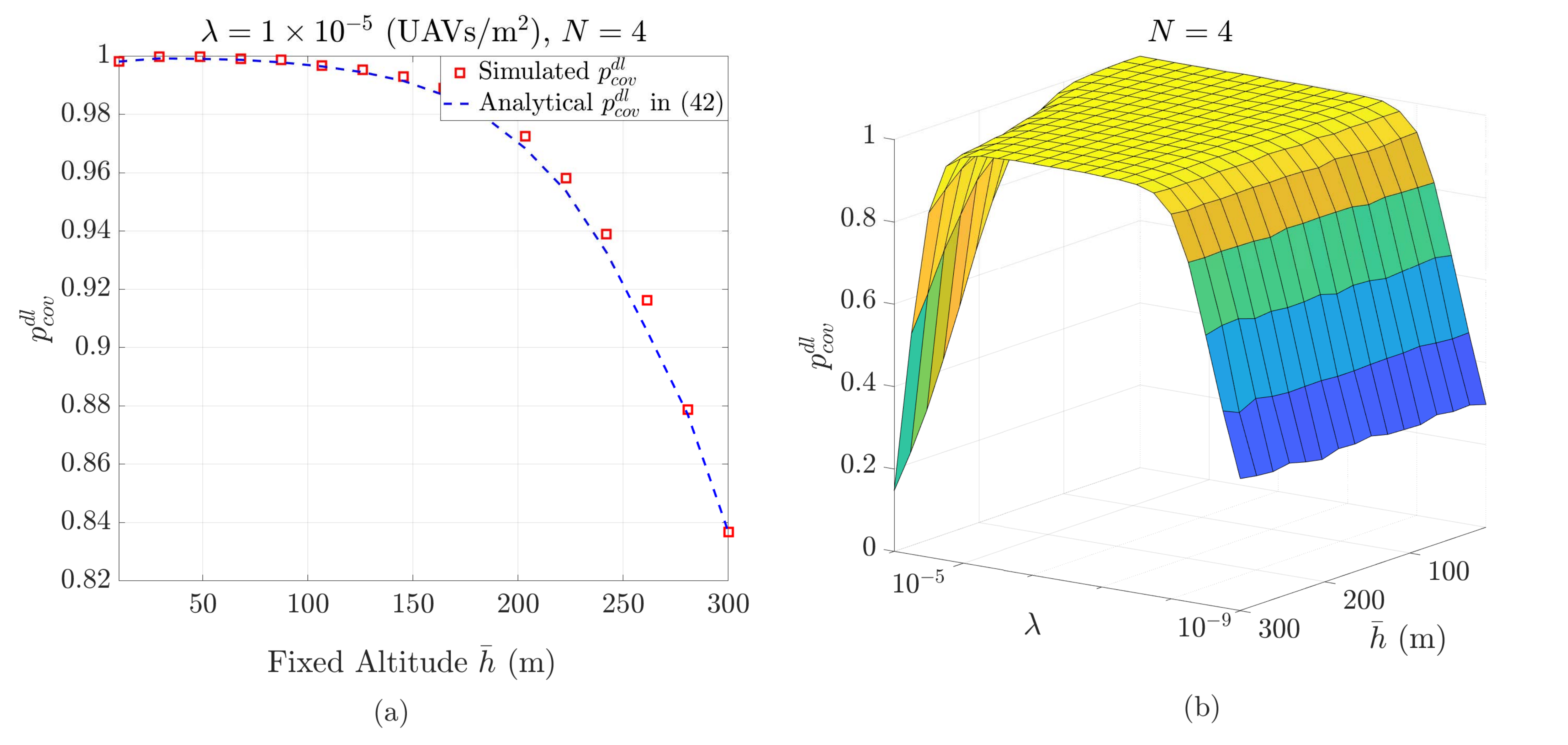}
	\caption{Simulation results of $p^{dl}_{cov}$ for the APDL scenario when $N=4$ and the altitude of each UAV is a constant $\overline{h}$ (i.e., $H=\bar{h}$): (a) The 2D simulation results of $p^{dl}_{cov}$ versus altitude $\overline{h}$ for $\lambda=1\times 10^{-5}$ (UAV/m$^2$), (b) The 3D simulation results of $p^{dl}_{cov}$ versus density $\lambda$ and the mean of altitude $\overline{h}$.}
	\label{Fig:APDL_FixedAltitude}
\end{figure*}

In this section, we will provide some numerical results to verify the previous analytical results of the downlink coverage. The numerical results of the APIL scenario will be presented and discussed first and those of the APDL will be shown and discussed afterwards. Finally, the numerical results of the downlink cell-free coverages will be presented. The network parameters adopted for simulation are shown in Table~\ref{Tab:SimPara} and they are chosen from the real statistical data provided in~\cite{AHKSSL14}. Other simulation parameters needed for the APIL and APDL scenarios will be specified in the following two subsections, respectively.

\subsection{Simulation Results for the APIL Scenario}\label{SubSection:SimAPIL}
In this subsection, we present the simulation results of the downlink coverage for the APIL scenario. Specifically, we consider the tangent of the elevation angle of a UAV is a Gamma RV with shape parameter $a$ and rate parameter $b$ (i.e., $\tan(\Theta)\sim\text{Gamma}(a,b)$) because using such a Gamma RV to model $\tan(\Theta)$ is able to generally characterize different distributions by setting different values of $a$ and $b$ so that appropriately adjusting $a$ and $b$ can make $\Theta$ reasonably distribute between $0$ and $\frac{\pi}{2}$. For example, $\tan(\Theta)$ becomes deterministic and equal to $\tan(\overline{\theta})$ such that $\Theta$ is equal to constant $\overline{\theta}$ if $b=a/\tan(\overline{\theta})$ and $a\rightarrow\infty$ and it becomes an exponential RV with rate parameter $1/b$ if $a=1$. Figures~\ref{Fig:APIL_ConstAng} and~\ref{Fig:APIL_GammaAng} show the simulation results of the downlink coverage $p^{dl}_{cov}$ when $\tan(\Theta)$ is a constant and a Gamma RV, respectively. As we can see, the simulation results of $p^{dl}_{cov}$ in Figs.~\ref{Fig:APIL_ConstAng}(a) and~\ref{Fig:APIL_GammaAng}(a) do not differ much when $\overline{\theta}<45^{\circ}$, which reveals that \textit{in general $p^{dl}_{cov}$ is insensitive to the distribution of $\Theta$} when the mean of $\Theta$ is not very large. In fact, this phenomenon can be inferred from~\eqref{Eqn:CovProbIndep} in that $p^{dl}_{cov}$ is affected by the distribution of $\Theta$ through $\omega$ in~\eqref{Eqn:Density3DPPInd} that is insensitive to the distribution of $\Theta$ when the mean of $\Theta$ is not large. Realizing this phenomenon is quite useful since we can quickly and accurately calculate $p^{dl}_{cov}$ using the mean of the elevation angle of UAVs in~\eqref{Eqn:CovProbIndep} without knowing the real distribution of $\Theta$, which is in general not easy to find in practice. 

Figures~\ref{Fig:APIL_ConstAng}(a) and~\ref{Fig:APIL_GammaAng}(a) validate the correctness and accuracy of the expression in~\eqref{Eqn:CovProbIndep} since the curve of the analytical result of $p^{dl}_{cov}$ in~\eqref{Eqn:CovProbIndep} completely coincides with the curve of the simulated result of $p^{dl}_{cov}$. Moreover, there exists an optimal value of the mean of $\Theta$ about $20^{\circ}$ for $\lambda=1\times 10^{-7}$ (UAVs/m$^2$), which maximizes $p^{dl}_{cov}$. Note that $p^{dl}_{cov}$ decreases as the mean of $\Theta$ increases over $20^{\circ}$ since the downlink SINR is now dominated by the interference in this situation even though the received signal power also increases. The 3D plots in Fig.~\ref{Fig:APIL_ConstAng}(b) and Fig.~\ref{Fig:APIL_GammaAng}(b) further show how $p^{dl}_{cov}$ varies with the mean of $\Theta$ and density $\lambda$. Generally speaking, the optimal value of the mean of $\Theta$ that maximizes $p^{dl}_{cov}$ changes with density $\lambda$ and $p^{dl}_{cov}$ converges up to a constant as $\lambda$ goes to infinity, i.e., $p^{dl}_{cov}$ barely depends on $\lambda$ as the network is dense and interference-limited, which is already shown in~\eqref{Eqn:CovProbIntLimit}.

\subsection{Simulation Results for the APDL Scenario}\label{SubSec:SimAPDL}

\begin{figure*}[t!]
	\centering
	\includegraphics[width=1\textwidth,height=3.0in]{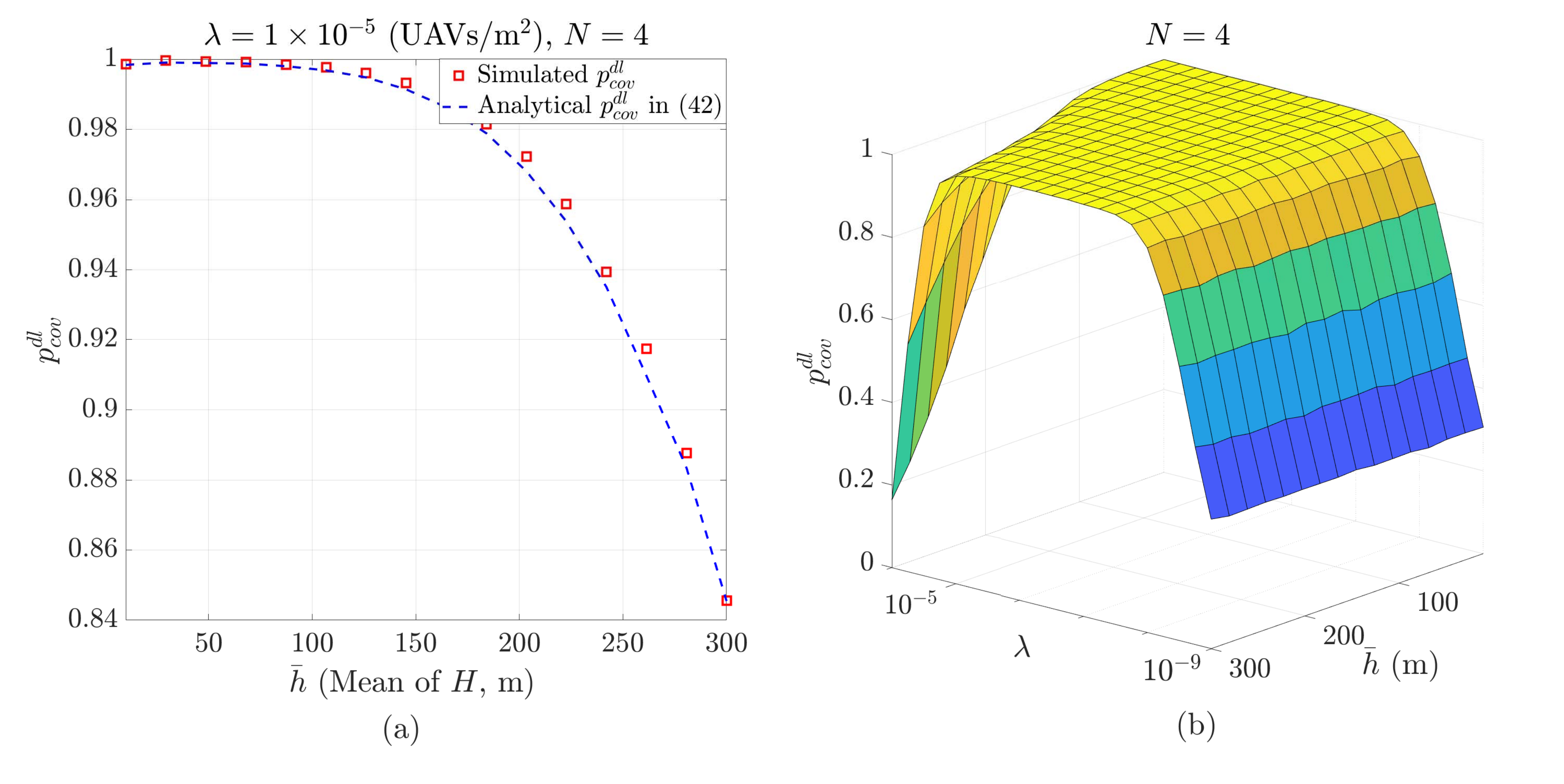}
	\caption{Simulation results of $p^{dl}_{cov}$ for the APDL scenario when $N=4$ and the altitude of each UAV is a uniformly distributed RV with mean $\overline{h}$ (i.e., $H\sim\text{Uni}[\overline{h}-a,\overline{h}+a]$ for $a>0$): (a) The 2D simulation results of $p^{dl}_{cov}$ versus altitude $\Theta$ for $\lambda=1\times 10^{-5}$ (UAV/m$^2$), (b) The 3D simulation results of $p^{dl}_{cov}$ versus density $\lambda$ for $H\sim\text{Uni}[\overline{h}-5, \overline{h}+5]$.}
	\label{Fig:APDL_UniAltitude}
\end{figure*}

\begin{figure*}[t!]
	\centering
	\includegraphics[width=1\textwidth,height=3.0in]{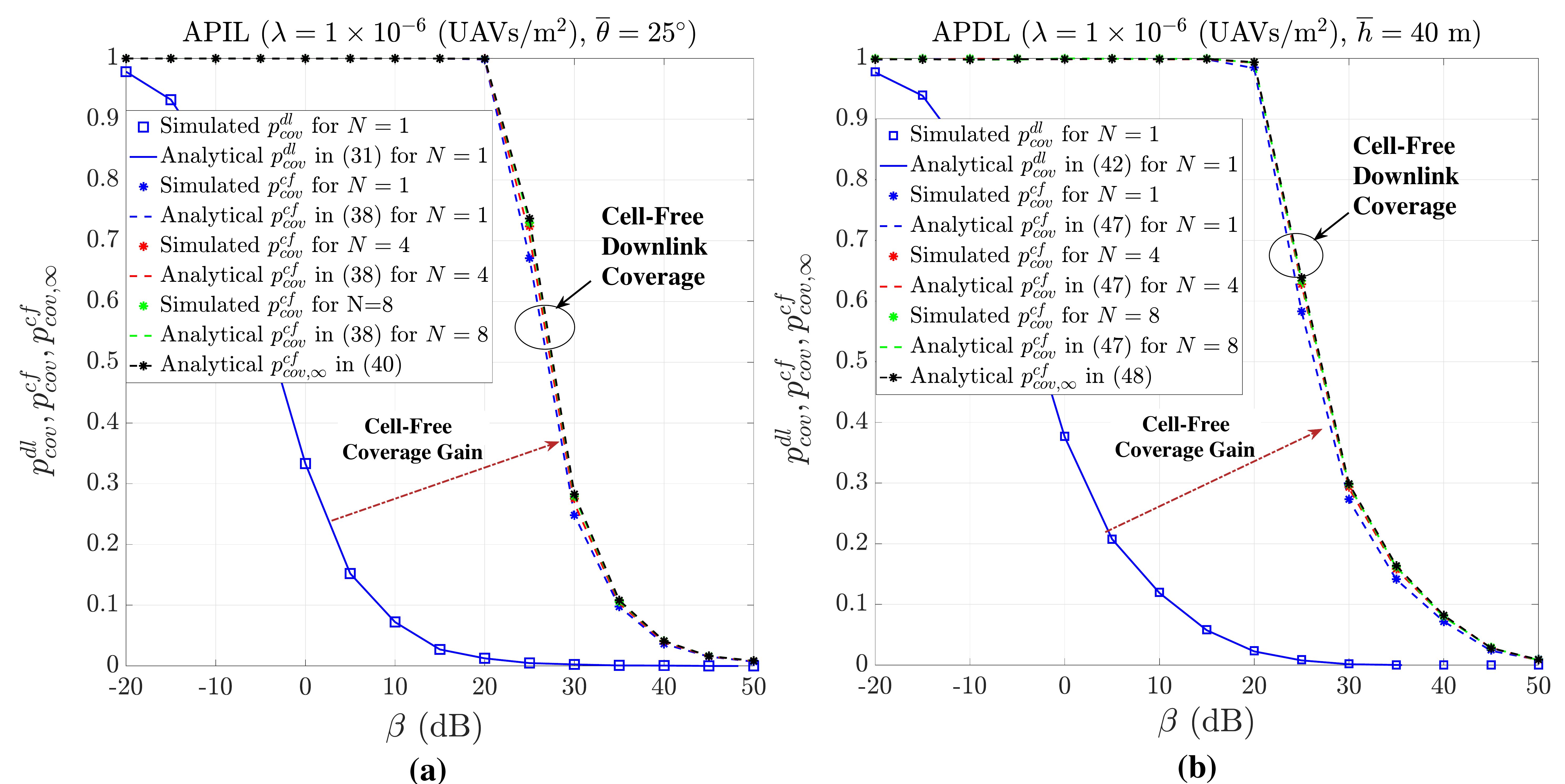}
	\caption{Simulation results of cell-free downlink coverage $p^{cf}_{cov}$ for $\lambda=1\times 10^{-6}$ (UAV/m$^2$) and $N=1, 2, 4, 8, \infty$: (a)  The simulation results of $p^{cf}_{cov}$ versus SINR threshold $\beta$ for the APIL scenario and the elevation angle of each UAV is a constant equal to $\overline{\theta}=5^{\circ}$, (b) The simulation results of $p^{cf}_{cov}$ versus SINR threshold $\beta$ for the APDL scenario and the altitude of each UAV is a constant equal to $\overline{h}=40$ m.}
	\label{Fig:CellFreeCovProb}
\end{figure*}

In this subsection, we specifically consider two distributions of the altitude of a UAV: one is \textit{deterministic} (fixed) altitude and the other is \textit{uniformly distributed} altitude. We would like to validate whether or not the analytical expression of $p^{dl}_{cov}$ in~\eqref{Eqn:CovProbAPDL} is correct and illustrate how $p^{dl}_{cov}$ varies with the two different distribution cases of the altitude of a UAV. When $N=4$, the simulation results of $p^{dl}_{cov}$ for the distribution case of fixed altitude and the distribution case of uniformly distributed altitude are shown in Figs.~\ref{Fig:APDL_FixedAltitude} and \ref{Fig:APDL_UniAltitude}, respectively. We see that the simulated results perfectly coincide with the analytical results of $p^{dl}_{cov}$ obtained from~\eqref{Eqn:CovProbAPDL} for $\lambda=1\times 10^{-5}$ (UAVs/m$^2$) in Figs.~\ref{Fig:APDL_FixedAltitude}(a) and \ref{Fig:APDL_UniAltitude}(a) so that the correctness of the expression in~\eqref{Eqn:CovProbAPDL} is validated. The simulation results in Figs.~\ref{Fig:APDL_FixedAltitude}(a) and \ref{Fig:APDL_UniAltitude}(a) are very close so that in general $p^{dl}_{cov}$ is insensitive to the different distributions of $H$ that have the same mean and thus $p^{dl}_{cov}$ can still be approximately calculated by~\eqref{Eqn:CovProbAPDL} with the mean of $H$ even when the real distribution $H$ is not known. Moreover, these two subplots both show that positioning UAVs too high significantly reduces $p^{dl}_{cov}$ thanks to LoS interference. It is noteworthy that $p^{dl}_{cov}$ degradation caused by LoS interference becomes apparent as $\lambda$ is high, which can be observed from Figs.~\ref{Fig:APDL_FixedAltitude}(b) and \ref{Fig:APDL_UniAltitude}(b), yet $p^{dl}_{cov}$ does not change much with the mean of $H$ due to low interference when $\lambda$ is small.

\subsection{Simulation Results for Cell-Free Downlink Coverage}

This subsection validates the analytical outcomes of the cell-free downlink coverage $p^{cf}_{cov}$ for the APIL and APDL scenarios. According to Fig.~\ref{Fig:CellFreeCovProb} that shows the numerical results of $p^{cf}_{cov}$, we can observe a few interesting and important phenomena.  First, the analytical results of the downlink cell-free coverages in both of the subplots perfectly coincide with their corresponding simulated results, which validates the correctness of the expressions in~ \eqref{Eqn:CellFreeCov1},~\eqref{Eqn:CellFreeCovInfN1},~\eqref{Eqn:CovProbAPDL},~\eqref{Eqn:CellFreeCovProbAPDL},~\eqref{Eqn:JfunAPDL1}, and~\eqref{Eqn:JfunAPDL2}. Second, the downlink cell-free coverages for different numbers of antennas are almost identical and this reveals that UAVs do not need to install multiple antennas to improve their coverage in the cell-free scenario so that UAVs can become lighter so as to save more power when flying. 
Third, the downlink cell-free coverage $p^{cf}_{cov}$ significantly outperforms the downlink coverage $p^{dl}_{cov}$, as can be seen in the figure. For example, $p^{cf}_{cov}$ for the APDL scenario and $\beta=0$ dB is able to achieve $100\%$, yet $p^{dl}_{cov}$ for the APDL scenario and $\beta=0$ dB is only about $12\%$.

\section{Conclusions}\label{Sec:Conclusion}

In the past decade, using 2D PPPs to model large-scale cellular networks had given rise to a great success in tractably analyzing the generic performance metrics of cellular networks. Nevertheless, straightforwardly employing a 3D PPP to deploy UAVs in a cellular network not only poses an unrealistic constraint on the path-loss exponent of 3D path-loss channel models, but also ignores a spatial deployment limitation in a cellular network, that is, in principle UAVs are low-altitude platforms that cannot be deployed in infinitely large 3D space modeled by a 3D PPP. Thus, there lack good 3D models with analytical tractability to deploy UAVs serving as aerial base stations in a large-scale cellular network. To tackle this issue, this paper proposes a 3D point process whose projections consist of a 2D homogeneous PPP and altitudes are the marks of the 2D homogeneous PPP. The fundamental properties of the proposed 3D point process are studied for the APIL and APDL scenarios and they pave a tractable way to analyze the downlink coverage of a UAV-enabled cellular network modeled by the proposed 3D point process. The downlink coverages for the APIL and APDL scenarios are explicitly derived and their closed-form expressions are also found for a special channel condition. In addition, cell-free downlink coverages and their upper limits are also derived when all the UAVs in the network can do non-coherence joint transmission.

\appendix [Proofs of Theorems and Propositions]
\numberwithin{equation}{section}
\setcounter{equation}{0}

\subsection{Proof of Theorem~\ref{Thm:CDFR*}}\label{App:ProofCDFR*}
(i) Consider the APIL scenario so that $X_i$ and $\Theta_j$ are independent for all $i,j\in\mathbb{N}_+$. Since $\|U_i\|=\|X_i\|\sec(\Theta_i)$, the CDF of $R_{\star}$ defined in~\eqref{Eqn:Weight3DDis} can be written as
\begin{align}
F_{R_{\star}}(r)&= \mathbb{P}\left[\max_{i:U_i\in\Phi_u} \left\{\frac{W_iL_i}{[\|X_i\|\sec(\Theta_i)]^{\alpha}}\right\}\leq r\right]\nonumber\\
&\stackrel{(a)}{=}\mathbb{E}\left\{\prod_{i:U_i\in\Phi_u}\mathbb{P}\left[\frac{W_iL_i}{[\|X_i\|\sec(\Theta_i)]^{\alpha}}\leq r\right]\right\}\nonumber
\end{align}
\begin{align}
\stackrel{(b)}{=}\exp\left(-2\pi\lambda\int_0^{\infty}\mathbb{P}\left[\frac{WL}{[x\sec(\Theta)]^{\alpha}}\geq r\right]x\dif x\right), \label{Eqn:AppCDFR*1}
\end{align}
where $(a)$ follows from the fact that all $W_iL_i[\|X_i\|\sec(\Theta_i)]^{-\alpha}$'s are independent and $(b)$ is obtained by first considering the independence between all RVs $W_i$, $L_i$, $\|X_i\|$, and $\Theta_i$ for all $i\in\mathbb{N}_+$ and then applying the probability generation functional (PGFL) of a homogeneous PPP to $\Phi_x$\footnote{Note that the subscript $i$ in $(a)$ is dropped in $(b)$ for notation simplification and such a subscript dropping is used throughout this paper whenever there is no notation ambiguity.}.  According to \eqref{Eqn:LoSProb}, $\mathbb{P}[WL[x\sec(\Theta)]^{-\alpha}\geq r|\Theta]$ can be further expressed as 
\begin{align*}
\mathbb{P}\left[\frac{WL}{[x\sec(\Theta)]^{\alpha}}\geq r \bigg|\Theta\right] =& \mathbb{P}\left[\left(\frac{W}{r}\right)^{\frac{1}{\alpha}}\cos(\Theta)\geq x\bigg|\Theta\right]\rho\left(\Theta\right)\\
&+\mathbb{P}\left[\left(\frac{\ell W}{r}\right)^{\frac{1}{\alpha}}\cos(\Theta)\geq x\bigg|\Theta\right]\\
&\times\left[1-\rho\left(\Theta\right)\right]. 
\end{align*}
Therefore, we can have the following:
\begin{align*}
&2\int_0^{\infty}\mathbb{P}\left[\frac{WL}{[x\sec(\Theta)]^{\alpha}}\geq r\bigg|\Theta\right]x\dif x \\
& = \rho(\Theta) \int_0^{\infty} \mathbb{P}\left[\left(\frac{W}{r}\right)^{\frac{1}{\alpha}}\cos(\Theta)\geq x\bigg|\Theta\right]\dif x^2+[1-\rho(\Theta)]\\
& \times \int_0^{\infty} \mathbb{P}\left[\left(\frac{\ell W}{r}\right)^{\frac{1}{\alpha}}\cos(\Theta)\geq x\bigg|\Theta\right]\dif x^2\\
&= \cos^2(\Theta)\left[\rho(\Theta) +[1-\rho(\Theta)]\ell^{\frac{2}{\alpha}}\right]\mathbb{E}\left[W^{\frac{2}{\alpha}}\right]r^{-\frac{2}{\alpha}}
\end{align*}
since $\int_0^{\infty}\mathbb{P}[Z\geq z]\dif z=\mathbb{E}[Z]$ for a non-negative RV $Z$. This gives rise to the following result:
\begin{align*}
& 2\int_0^{\infty}\mathbb{P}\left[\frac{WL}{[x\sec(\Theta)]^{\alpha}}\geq r\right]x\dif x=\mathbb{E}\left[W^{\frac{2}{\alpha}}\right] r^{-\frac{2}{\alpha}}\\
&\times\mathbb{E}\left\{\cos^2(\Theta)\left[\rho(\Theta) \left(1-\ell^{\frac{2}{\alpha}}\right)+\ell^{\frac{2}{\alpha}}\right]\right\},
\end{align*}
and then substituting this identity into~\eqref{Eqn:AppCDFR*1} yields the expression in~\eqref{Eqn:CDFR*1}.

(ii) Consider the APDL scenario and we know $\Theta_i=\tan^{-1}(H_i/\|X_i\|)$ for $H_i>0$. Since we know $\sec^2(\Theta_i)=1+H^2_i/\|X_i\|^2$ and $\Theta=\tan^{-1}(H/x)$, $\mathbb{P}[WL[x\sec(\Theta)]^{-\alpha}\geq r]$ in \eqref{Eqn:AppCDFR*1} for a given $H$ can be rewritten as
\begin{align*}
&\mathbb{P}\left[\frac{WL}{x^{\alpha}(1+H^2/x^2)^{\frac{\alpha}{2}}}\geq r \bigg| H\right] =\mathbb{P}\left[W\geq r(x^2+H^2)^{\frac{\alpha}{2}}|H\right]\\
&\times\rho\left(\Theta\right)+\left[1-\rho\left(\Theta\right)\right] \mathbb{P}\left[W\geq \frac{r}{\ell}  (x^2+H^2)^{\frac{\alpha}{2}}|H\right]=\rho\left(\Theta\right) \\
&\times\mathbb{P}\left[W\geq r x^{\alpha}\sec^{\alpha}(\Theta)|H\right]+[1-\rho(\Theta)]\\ &\times\mathbb{P}\left[W\geq \frac{r}{\ell} x^{\alpha}\sec^{\alpha}(\Theta)|H\right].
\end{align*}
Substituting this identity into~\eqref{Eqn:AppCDFR*1} and replacing $x^2$ with $z$ yield the expression in~\eqref{Eqn:CDFR*2}.

\subsection{Proof of Theorem~\ref{Thm:LapTran1}}\label{App:ProofLapTran1}
First, consider the APIL scenario in which the elevation angle and projection of point $U_i$ are independent. Since the projections of all the points in $\Phi_u$ is a 2D homogeneous PPP of density $\lambda$ and $\|U_i\|=\|X_i\|\sec(\Theta_i)$, the Laplace transform, $\mathcal{L}_{T_0}(s)=\mathbb{E}[\exp(-sT_0)]$, can be found as follows:
\begin{align}
&\mathbb{E}\left[e^{-sT_0}\right]=\mathbb{E}\left[\prod_{i:U_i\in\Phi_u}\exp\left(-\frac{sW_iL_i}{\|U_i\|^{\alpha}}\right)\right]\nonumber\\
&= \mathbb{E}_{\Phi_u}\left\{\prod_{i:X_k\in\Phi_x} \mathbb{E}_{WL}\left[\exp\left(-\frac{sW_iL_i}{(\|X_i\|\sec(\Theta_i))^{\alpha}}\right)\right]\right\}\nonumber\\
&\stackrel{(a)}{=}\exp\left(-\pi \lambda\int_0^{\infty}\left\{1-\mathbb{E}\left[e^{-sWL(x\sec(\Theta))^{-\alpha}}\right]\right\}\dif x^2 \right) \label{Eqn:ProofLapTransCompSSP0}\\
&\stackrel{(b)}{=}\exp\left(-\pi\lambda\int_0^{\infty}\mathbb{P}\left[Y\leq sWL\cos^{\alpha}(\Theta)z^{-\frac{\alpha}{2}}\right]\dif z\right)\nonumber\\
&= \exp\left(-\pi\lambda\int_0^{\infty} \mathbb{P}\left[z \leq \cos^2(\Theta)\left(\frac{sWL}{Y}\right)^{\frac{2}{\alpha}}\right] \dif z\right),\label{Eqn:ProofLapTransCompSSP1}
\end{align}
where $(a)$ is obtained by applying the PGFL of a homogeneous PPP to $\Phi_x$ and $(b)$ is obtained by first replacing $x^2$ with $z$ and then rewriting the result in the integral by using $Y\sim\exp(1)$). In addition, we can have
\begin{align*}
&\int_0^{\infty}\mathbb{P}\left[z \leq \cos^2(\Theta)\left(\frac{sWL}{Y}\right)^{\frac{2}{\alpha}}\right] \dif z = s^{\frac{2}{\alpha}} \mathbb{E}\left[W^{\frac{2}{\alpha}}\right]\mathbb{E}\left[Y^{-\frac{2}{\alpha}}\right]\\
&\mathbb{E}\left[\cos^2(\Theta)L^{\frac{2}{\alpha}}\right]\stackrel{(c)}{=} s^{\frac{2}{\alpha}}\Gamma\left(1-\frac{2}{\alpha}\right)\mathbb{E}\left[W^{\frac{2}{\alpha}}\right]\mathbb{E}\bigg[\cos^2(\Theta)\\
&\times\left(\rho(\Theta)\left(1-\ell^{\frac{2}{\alpha}}\right)+\ell^{\frac{2}{\alpha}}\right)\bigg],
\end{align*}
where $(c)$ is acquired by using the two facts that $\mathbb{E}[Y^{-\frac{2}{\alpha}}]=\Gamma(1-\frac{2}{\alpha})$ and $\mathbb{E}[\cos^2(\Theta)L^{\frac{2}{\alpha}}|\Theta]=\cos^2(\Theta)(\rho(\Theta)+[1-\rho(\Theta)]\ell^{\frac{2}{\alpha}})$ for a given $\Theta$. Substituting this result into~\eqref{Eqn:ProofLapTransCompSSP1} yields the expression in~\eqref{Eqn:LapTranComSSP1}.

Next, consider the APDL scenario so that $\Theta_i=\tan^{-1}(H_i/\|X_i\|)$ for all $i\in\mathbb{N}_+$. For this scenario, the integral in~\eqref{Eqn:ProofLapTransCompSSP1} for a given $H$ can be expressed as
\begin{align}
& \mathbb{P}\left[z\leq\cos^2(\Theta) \left(\frac{sWL}{Y}\right)^{\frac{2}{\alpha}}\bigg| H\right] =\mathbb{P}\left[Y\leq \frac{sW}{z^{\frac{\alpha}{2}}\sec^{\alpha}(\Theta)}\bigg|H\right]\nonumber\\
&\rho\left(\Theta\right)+\mathbb{P}\left[Y\leq \frac{s\ell W}{z^{\frac{\alpha}{2}}\sec^{\alpha}(\Theta)}\bigg|H\right][1-\rho(\Theta)]=\rho(\Theta)\nonumber\\
&\times\bigg[\mathcal{L}_W\left(z^{-\frac{\alpha}{2}}s\ell\cos^{\alpha}(\Theta)\right)-\mathcal{L}_W\left(z^{-\frac{\alpha}{2}}s\cos^{\alpha}(\Theta)\right)\bigg]\nonumber\\
&+1-\mathcal{L}_W\left(z^{-\frac{\alpha}{2}}s\ell\cos^{\alpha}(\Theta)\right).  \label{Eqn:ProofLapTranComSSP2}
\end{align}
We can get the expression in~\eqref{Eqn:LapTranComSSP2} by substituting~\eqref{Eqn:ProofLapTranComSSP2} into~\eqref{Eqn:ProofLapTransCompSSP1}.

\subsection{Proof of Theorem~\ref{Thm:LapTran2}}\label{App:ProofLapTran2}

In Section~\ref{SubSec:Dis-RelDis}, we have pointed out that $\Phi_u$ is equivalently equal to a 2D homogeneous PPP of density $\lambda_u=\lambda\mathbb{E}[\cos^2(\Theta)]$ in the APIL scenario. Let $U_K$ denote the $K$th nearest point in $\Phi_u$ to the origin and its projection is $X_K$. As such, the CCDF of $\|U_K\|^2$ can be expressed as follows~\cite{MH12,DSWKJM13,CHLLCW16}:
\begin{align*}
\mathbb{P}\left[\|U_K\|^2 \geq x\right] &=  \sum_{k=0}^{K-1}\frac{(\pi\lambda_u x)^k}{k!}e^{-\pi\lambda_u u}\\
&=\mathbb{P}\left[\|X_K\|^2 \sec^2(\Theta_K)\geq x\right],
\end{align*}
where $\Theta_K$ is the elevation angle of $U_K$. Thus, the CCDF of $\|U_K\|^2$ reduces to the CCDF of $\|X_K\|^2$ whenever $\Theta_K=0$ for any $K$. This follows that
\begin{align*}
\mathbb{P}\left[\|X_K\|^2\geq x\cos^2(\Theta_K)|\Theta_K=0\right] &= \mathbb{P}\left[\|X_K\|^2\geq x\right]\\
&= \sum_{k=0}^{K-1}\frac{(\pi\lambda x)^k}{k!}e^{-\pi\lambda x},
\end{align*}
which indicates $\|X_K\|^2\sim\text{Gamma}(K,\pi\lambda)$. For the APDL scenario, $\Phi_u$ can be equivalently equal to a 2D non-homogeneous PPP of density $\lambda F_H(\sqrt{z})$, as already shown in Section~\ref{SubSec:Dis-RelDis}. As a result, the CCDF of $\|U_K\|^2$ in this scenario can be written as
\begin{align*}
\mathbb{P}\left[\|U_K\|^2\geq x\right] = \sum_{k=0}^{K-1}\frac{[\pi \int_0^{x}\lambda F_H(\sqrt{z})\dif z]^k}{e^{\pi \int^{x}_0 \lambda F_H(\sqrt{z})\dif z}k!}.
\end{align*}
Setting $H=0$ in the above result shows $\|X_K\|^2\sim\text{Gamma}(K,\pi\lambda)$. These above results manifest that the point ordering in $\Phi_u$ is the same as that in of the projections of $\Phi_u$ that are a homogeneous PPP of density $\lambda_x$ no matter whether or not the elevation angle and the projection of each point in $\Phi_u$ are independent.  

Now consider the APIL scenario in which $\Theta_k$ and $X_k$ of $U_k\in\Phi_u$ are independent for all $k\in\mathbb{N}_+$. Since $U_k$ is the $k$th nearest point in $\Phi_u$ that is a homogeneous PPP, we know $\|U_{i+K}\|^2=\|U_K\|^2+\|U_i\|^2$ where $\|U_K\|$ and $\|U_i\|$ are independent~\cite{CHLLCW16}, $\mathcal{L}_{T_K}(s)$ can be explicitly expressed as shown in the following:
\begin{align}
&\mathcal{L}_{T_K}(s) = \mathbb{E}\left\{\exp\left(-s\sum_{i=1}^{\infty} \frac{W_{i+K}L_{i+K}}{\|U_{i+K}\|^{\alpha}}\right)\right\}\nonumber\\
=&\mathbb{E}\left\{\exp\left(-s \sum_{i=1}^{\infty} \dfrac{W_{i+K}L_{i+K}\cos^{\alpha}(\Theta_{i+K})}{\|X_{K+i}\|^{\alpha}}\right)\right\}\nonumber\\
 \stackrel{(a)}{=}&\mathbb{E}\left\{\mathbb{E}\left[\prod_{i:X_i\in\Phi_x}\exp\left(-\dfrac{sW_iL_i\cos^{\alpha}(\Theta_{i})}{\left(\|X_K\|^2+\|X_i\|^2\right)^{\frac{\alpha}{2}}} \right)\bigg|\|X_K\|\right]\right\}\nonumber\\
\stackrel{(b)}{=}&  \mathbb{E}\bigg\{\exp\bigg[-\pi\lambda\int_0^{\infty} \bigg(1-\mathbb{E}\bigg[\exp\left(- \dfrac{sWL\cos^{\alpha}(\Theta)}{\left(\|X_K\|^2+x\right)^{\frac{\alpha}{2}}}\right) \nonumber\\ 
&\bigg| \|X_K\|\bigg]\bigg)\dif x\bigg]\bigg\}\label{Eqn:ProofLapTranTrunSSP0}\\
\stackrel{(c)}{=} &\mathbb{E}\bigg\{\exp\bigg(-\pi\lambda D_K \int_{1}^{\infty} \bigg\{1-\mathbb{E}_{L,\Theta}\left[\mathcal{L}_{W}\left(\frac{sL\cos^{\alpha}(\Theta)}{z^{\frac{\alpha}{2}}D_K^{\frac{\alpha}{2}}}\right)\right]\nonumber\\ &\bigg\}\dif z \bigg)\bigg\},\label{Eqn:ProofLapTranTrunSSP1}
\end{align}
where $(a)$ follows from the fact that all $L_i$'s ($W_i$'s) are i.i.d. and $X_K$ ($X_i$) is the projection of $U_K$ ($U_i$), $(b)$ is obtained by the PGFL of a 2D homogeneous PPP to $\Phi_x$, and $(c)$ is obtained by first replacing $\|X_K\|^2+x$ with $\|X_K\|^2 z$ in the integral and replacing $\|X_K\|^2$ with $D_K$. Also, the integral in~\eqref{Eqn:ProofLapTranTrunSSP1} can be simplified as shown in the following:
\begin{align*}
 &\int_1^{\infty}\left\{1-\mathbb{E}_{L,\Theta}\left[\mathcal{L}_{W}\left(\frac{sL\cos^{\alpha}(\Theta)}{z^{\frac{\alpha}{2}}D_K^{\frac{\alpha}{2}}}\right)\right] \right\}\dif z \\
=& \mathbb{E}_{\Theta}\bigg\{\rho(\Theta) \int_1^{\infty}\left[1-\mathcal{L}_{W}\left(\frac{s\cos^{\alpha}(\Theta)}{(zD_K)^{\frac{\alpha}{2}}}\right)\right]\dif z\\
&+[1-\rho(\Theta)] \int_1^{\infty}\left[1-\mathcal{L}_{W}\left(\frac{s\ell\cos^{\alpha}(\Theta)}{(zD_K)^{\frac{\alpha}{2}}}\right)\right]\dif z\bigg\},
\end{align*}
and we can further show
\begin{align*}
\int_1^{\infty} \left[1-\mathcal{L}_W\left(xz^{-\frac{\alpha}{2}}\right)\right]\dif z=\mathcal{I}_W\left(x,\frac{2}{\alpha}\right)
\end{align*}
by following the derivation techniques in the proof of Proposition 1 in~\cite{CHLLCW1502} and using the definition of $\mathcal{I}_W(x,y)$ in~\eqref{Eqn:Ifun}. Thus, we finally get the result in~\eqref{Eqn:LapTranImSSP1} owing to $D_K\sim\text{Gamma}(K,\pi\lambda)$.
 

Now consider the APDL scenario such that $\Theta_i=\tan^{-1}(H_i/\|X_i\|)$. From~\eqref{Eqn:ProofLapTranTrunSSP0},we can rewrite the expression inside the integral with $\Theta=\tan^{-1}(H/\sqrt{\|X_K\|^2+x})$ and $\|X_K\|^2+x=D_K+x=z$ as follows:
\begin{align*}
& 1-\mathbb{E}_L\left[\mathcal{L}_{W}\left(\frac{sL\cos^{\alpha}(\Theta)}{z^{\frac{\alpha}{2}}}\right)\bigg|H\right]=1-\rho(\Theta)\\
& \times\mathcal{L}_{W}\left(\frac{s\cos^{\alpha}(\Theta)}{z^{\frac{\alpha}{2}}}\right)-[1-\rho(\Theta)]\mathcal{L}_{W}\left(\frac{s\ell\cos^{\alpha}(\Theta)}{z^{\frac{\alpha}{2}}}\right)\\ 
&= \mathcal{J}_W\left(\frac{s}{z^{\frac{\alpha}{2}}}, \tan^{-1}\left(\frac{H}{\sqrt{z}}\right)\right).
\end{align*}
Substituting this above result into~\eqref{Eqn:ProofLapTranTrunSSP1} and then averaging the whole expression over $D_K\sim\text{Gamma}(K,\pi\lambda)$ lead to~\eqref{Eqn:LapTranImSSP2}. This completes the proof.

\subsection{Proof of Proposition~\ref{Prop:CovProbIndep}}\label{App:ProofCovProbIndep}

For the APIL scenario, we can infer the following from \eqref{Eqn:AssoUAV}:
\begin{align*}
\min_{i:U_i\in\Phi_u}\{L^{-\frac{1}{\alpha}}_i\|U_i\|\}\stackrel{d}{=}\min_{i:\widetilde{U}_i\in\widetilde{\Phi}_u}\|\widetilde{U}_i\|\defn\|\widetilde{U}_{\star}\|,
\end{align*}
where $\widetilde{\Phi}_u$ already defined in~\eqref{Eqn:Equ3DPointProcess} is a homogeneous PPP of density $\lambda\omega$ as shown in Section~\ref{SubSec:Dis-RelDis} and $\widetilde{U}_{\star}$ is the nearest point in $\widetilde{\Phi}_u$ to the origin.  Thus, we know
\begin{align*}
\widetilde{I}_0 = \sum_{i:\widetilde{U}_i\in\widetilde{\Phi}_u\setminus \widetilde{U}_{\star}} P G_i\|\widetilde{U}_i\|^{-\alpha}\stackrel{d}{=} I_0.
\end{align*}
The CCDF of a non-negative RV $Z$ can be expressed as
\begin{align}\label{Eqn:CCDFIden1}
F^c_Z(z) = \mathcal{L}^{-1}\left\{\frac{1}{s}\mathcal{L}_{Z^{-1}}\left(s\right)\right\}\left(\frac{1}{z}\right),\,\, s>0.
\end{align}
It can be used to express $p_{cov}$ in~\eqref{Eqn:KjointCovProb} for $G_i\sim\exp(1)$ as follows:
\begin{align}
p_{cov} =&\mathcal{L}^{-1}\left\{\frac{1}{s}\mathcal{L}_{\gamma^{-1}_0}(s)\right\}\nonumber\\
=&\mathcal{L}^{-1}\left\{\mathbb{E}\left[\frac{1}{s}\exp\left(-s\frac{(\widetilde{I}_0+\sigma_0)\|U_{\star}\|^{\alpha}}{PG_{\star}L_{\star}}\right)\right]\right\}\left(\frac{1}{\beta}\right)\nonumber\\
=& \mathcal{L}^{-1}\bigg\{ \mathbb{E}\bigg[\frac{1}{s}\exp\left(- \frac{s\sigma_0\|\widetilde{U}_{\star}\|^{\alpha}}{PG_{\star}}\right)\nonumber\\
&\times\mathcal{L}_{\widetilde{I}_0|\|\widetilde{U}_{\star}\|}\left(\frac{s\|\widetilde{U}_{\star}\|^{\alpha}}{PG_{\star}}\right)\bigg]\bigg\}\left(\frac{1}{\beta}\right), \label{Eqn:ProofCovProbIndeCase}
\end{align}
where $\mathcal{L}_{\widetilde{I}_0|\|\widetilde{U}_{\star}\|}\left(\cdot\right)$ is the Laplace transform of $\widetilde{I}_0$ while conditioning on $\|\widetilde{U}_{\star}\|$. Note that $\widetilde{I}_0$ is the first-truncated shot signal process in $\widetilde{\Phi}_u$ since the projection of point $\widetilde{U}_{\star}$ is the nearest point among all the projections of the points in $\widetilde{\Phi}_u$, that is, $\widetilde{I}_0$ is equal to $T_K$ in~\eqref{Eqn:TrucKthSSS} for $K=1$, $W_k=PG_k$, and $L_k=1$.    

Since $\widetilde{I}_0$ is the first-truncated shot signal process in $\widetilde{\Phi}_u$,  $W=PG$, and $\|\widetilde{U}_{\star}\|^2=D_{\star}\sim\exp(\pi\lambda\omega)$, $\mathcal{L}_{\widetilde{I}_0|\|\widetilde{U}_{\star}\|}\left(\cdot\right)$ in~\eqref{Eqn:ProofCovProbIndeCase} can be found by using $K=1$ and replacing $s$ with $s\|\widetilde{U}_{\star}\|^{\alpha}/PG_{\star}$ in~\eqref{Eqn:LapTranImSSP1} as follows
\begin{align*}
\mathcal{L}_{\widetilde{I}_0|\|\widetilde{U}_{\star}\|}\left(\frac{s\|\widetilde{U}_{\star}\|^{\alpha}}{PG_{\star}}\right) &=\exp\left[-\pi\lambda\omega D_{\star} \mathcal{I}_W\left(\frac{s}{PG_{\star}},\frac{2}{\alpha}\right)\right]\\
&=\exp\left[-\pi\lambda\omega D_{\star} \mathcal{I}_G\left(\frac{s}{G_{\star}},\frac{2}{\alpha}\right)\right],
\end{align*}
where $\mathcal{I}_W(s/PG_{\star},2/\alpha)$ for $W=PG$ and $G\sim\exp(1)$ is equal to $\mathcal{I}_G(s/G_{\star},2/\alpha)$ defined in~\eqref{Eqn:IfunNoW}.  Then substituting this result into~\eqref{Eqn:ProofCovProbIndeCase} yields 
\begin{align}
p_{cov} = & \mathcal{L}^{-1}\bigg\{\mathbb{E}\bigg[\frac{1}{s}\exp\bigg(- \frac{s\sigma_0D_{\star}^{\frac{\alpha}{2}}}{PG}-\pi\lambda \omega D_{\star}\nonumber\\
&\times\mathcal{I}_G\left( \frac{s}{G_{\star}},\frac{2}{\alpha}\right)\bigg) \bigg]\bigg\}\left(\frac{1}{\beta}\right). \label{Eqn:ProofCovProbIndeCase2}
\end{align}
Furthermore, we know the following identity for a real-valued  function $\psi:\mathbb{R}_+\rightarrow \mathbb{R}_+$:
\begin{align}\label{Eqn:LapTransIden1}
\mathbb{E}\left\{\frac{1}{s}\exp\left[-\psi\left(\frac{s}{Z}\right)\right] \right\}=\int_0^{\infty} \exp\left[-\psi\left(\frac{1}{t}\right)\right]f_Z(st)\dif t,
\end{align}
where $f_Z(z)$ is the PDF of $Z$. This follows that
\begin{align}
&\mathbb{E}\left[\frac{1}{s}\exp\left(- \frac{s\sigma_0D_{\star}^{\frac{\alpha}{2}}}{PG_{\star}}-\pi\lambda \omega D_{\star}\mathcal{I}_G\left( \frac{s}{G_{\star}},\frac{2}{\alpha}\right)\right) \right]=\nonumber\\ &\int_0^{\infty} \exp\left[-\frac{\sigma_0D^{\frac{\alpha}{2}}_{\star}}{Pt}-\pi\lambda\omega D_{\star}\mathcal{I}_G\left(\frac{1}{t},\frac{2}{\alpha}\right)\right]f_{G_{\star}}(st)\dif t\nonumber\\ &\stackrel{(a)}{=}\frac{1}{(N-1)!}\int_0^{\infty}\exp\left[-\frac{\sigma_0D^{\frac{\alpha}{2}}_{\star}}{Pt}-\pi\lambda\omega D_{\star}\mathcal{I}_G\left(\frac{1}{t},\frac{2}{\alpha}\right)\right]\nonumber\\
&(st)^{N-1}e^{-st} \dif t\stackrel{(b)}{=}\frac{s^{N-1}}{(N-1)!}\int_0^{\infty} \exp\bigg[-\frac{\sigma_0 D^{\frac{\alpha}{2}}_{\star}}{P t}-\pi\lambda\omega  
\end{align}
\begin{align}
&\times D_{\star}\mathcal{I}_G\left(\frac{1}{t},\frac{2}{\alpha}\right)\bigg]t^{N-1} e^{-st} \dif t=\mathcal{L}\bigg\{\frac{1}{(N-1)!}\frac{\dif^{N-1}}{\dif t^{N-1}}\nonumber\\
&\bigg[t^{N-1}\exp\left(-\frac{\sigma_0 D^{\frac{\alpha}{2}}_{\star}}{P t}-\pi\lambda\omega D_{\star}\mathcal{I}_G\left(\frac{1}{t},\frac{2}{\alpha}\right)\right)\bigg]\bigg\}(s), \label{Eqn:ProofCovProbIndeCase3}
\end{align}
where $(a)$ is obtained due to $G_{\star}\sim\text{Gamma}(N,1)$ and $(b)$ is obtained by moving $s^{N-1}$ out of the integral.  We then substitute~\eqref{Eqn:ProofCovProbIndeCase3} into \eqref{Eqn:ProofCovProbIndeCase2} to get $p_{cov}$ as shown in~\eqref{Eqn:CovProbIndep}.  
 
\subsection{Proof of Proposition~\ref{Prop:CellFreeCovProbAPIL}}\label{App:ProofCellFreeCoverage} 
By letting $S_0\defn \sum_{i:U_i\in\Phi_u}G_iL_i\|U_i\|^{-\alpha}$ and using~\eqref{Eqn:CCDFIden1}, the cell-free downlink coverage defined in~\eqref{Eqn:DefnCellFreeCovProb} can be rewritten as
\begin{align}
p^{cf}_{cov} &= 1-\mathbb{P}\left[S_0\leq \frac{\beta\sigma_0}{P}\right]=1-F^c_{S^{-1}_0}\left(\frac{P}{\beta\sigma_0}\right)\nonumber\\
&=1-\mathcal{L}^{-1}\left\{\frac{1}{s}\mathcal{L}_{S_0}\left(s\right)\right\}\left(\frac{\beta\sigma_0}{P}\right). \label{Eqn:ProofCellFreeCov}
\end{align}
Note that $S_0\stackrel{d}{=}T_0$ defined in~\eqref{Eqn:ComSSP} with $G_i=W_i$ so that $S_0$ is a Poisson shot signal process in $\Phi_u$. 
According to Theorem~\ref{Thm:LapTran1}, $\mathcal{L}_{S_0}(s)=\mathcal{L}_{I_0}(s)$ for $W\sim\text{Gamma}(N,1)$ can be found as
\begin{align*}
\mathcal{L}_{S_0}(s) &=  \exp\left\{-\pi\lambda s^{\frac{2}{\alpha}}\mathbb{E}\left[W^{\frac{2}{\alpha}}\right]\Gamma\left(1-\frac{2}{\alpha}\right)\omega\right\}\nonumber\\
&=\exp\left[-\frac{\pi\lambda s^{\frac{2}{\alpha}}\omega}{(N-1)!}\Gamma\left(N+\frac{2}{\alpha}\right)\Gamma\left(1-\frac{2}{\alpha}\right)\right].
\end{align*}
Substituting this into~\eqref{Eqn:ProofCellFreeCov} leads to~\eqref{Eqn:CellFreeCov1}. For $\alpha=4$, \eqref{Eqn:CellFreeCov1} further reduces to~\eqref{Eqn:CellFreeCov2} since the inverse Laplace transform can be found in closed form~\cite{MAIA12}. 

\subsection{Proof of Proposition~\ref{Prop:CovProbDep}}\label{App:ProofCovProbDep}
Since the APDL scenario is considered, we know $\widetilde{\Phi}_u$ is a 2D non-homogeneous PPP of density $\lambda\frac{\dif\Omega(y^{\frac{\alpha}{2}})}{\dif y}$ with $\Omega(y^{\alpha/2})$ given in~\eqref{Eqn:OmegaFunLos}, as shown in Section~\ref{SubSec:Dis-RelDis}. Moreover, $\widetilde{I}_0$ is the first-truncated shot signal process in $\widetilde{\Phi}_u$, as pointed out in Appendix~\ref{App:ProofCellFreeCoverage}. We thus are able to express $\mathcal{L}_{\widetilde{I}_0|\|\widetilde{U}_{\star}\|}(\cdot)$ in~\eqref{Eqn:ProofCovProbIndeCase} by using $K=1$, $W=PG$, $G\sim\exp(1)$, $L=1$, and replacing $D_K$ with $\|\widetilde{U}_{\star}\|^2=\widetilde{D}_{\star}$ in~\eqref{Eqn:LapTranImSSP2} as follows:
\begin{align}
&\mathcal{L}_{\widetilde{I}_0|\|\widetilde{U}_{\star}\|}\left(\frac{s\|\widetilde{U}_{\star}\|^{\alpha}}{PG_{\star}}\right)\nonumber\\ 
&=\mathbb{E}\left[\exp\left(-\frac{s\widetilde{D}_{\star}^{\frac{\alpha}{2}}}{G_{\star}} \sum_{i:\widetilde{U}_i\in\widetilde{\Phi}_u\setminus \widetilde{U}_{\star}} \frac{G_i}{(\widetilde{D}_{\star}+\|\widetilde{U}_i\|^2)^{\frac{\alpha}{2}}}\right)\right]\nonumber \\
&\stackrel{(a)}{=}\exp\left[-\pi\lambda\int_{\widetilde{D}_{\star} }^{\infty}\left[\frac{\dif \Omega\left(y^{-\frac{\alpha}{2}}\right)}{\dif y}\right]\mathcal{J}_W\left(\dfrac{s \widetilde{D}^{\frac{\alpha}{2}}_{\star}}{G_{\star}y^{\frac{\alpha}{2}}},0 \right)\dif y\right]\nonumber\\
&\stackrel{(b)}{=}\exp\left[-\pi \lambda \widetilde{D}_{\star} \int_1^{\infty} \left(\frac{s}{G_{\star}y^{\frac{\alpha}{2}}+s}\right) \dif\Omega\left(\frac{\widetilde{D}_{\star}}{y^{\frac{\alpha}{2}}}\right)\right],\label{Eqn:ProofLapAPDL}
\end{align}
where $(a)$ is obtained by applying the result in~\eqref{Eqn:LapTranImSSP2} for $\widetilde{\Phi}_u$ and $K=1$ and then assuming $\widetilde{\Theta}_{\star}=\tan^{-1}\frac{H}{\sqrt{y\widetilde{D}_{\star}}}$ is the elevation angle between the typical user and $\widetilde{U}_{\star}$, and $(b)$ is because $\mathcal{J}_W(\cdot,\cdot)$ in~\eqref{Eqn:WfunIndep} for $W=G$ and $\ell=1$ can be expressed as
\begin{align*}
\mathcal{J}_W\left(\dfrac{s}{G_{\star}y^{\frac{\alpha}{2}}},0 \right)= 1-\mathcal{L}_G\left(\frac{s}{G_{\star}y^{\frac{\alpha}{2}}}\right)=\frac{s}{G_{\star}y^{\frac{\alpha}{2}}+s}.
\end{align*}
Substituting \eqref{Eqn:ProofLapAPDL} into~\eqref{Eqn:ProofCovProbIndeCase} and following the steps of deriving~\eqref{Eqn:ProofCovProbIndeCase3} yield the result in~\eqref{Eqn:CovProbAPDL}.


\bibliographystyle{IEEEtran}
\bibliography{IEEEabrv,Ref_3DUAVCellular}

\end{document}